\documentclass[notitlepage,aps,preprintnumbers,onecolumn,amsmath,amssymb,floatfix,pra]{revtex4-1}

\pdfoutput=1
\usepackage[utf8]{inputenc}

\usepackage{amsmath}
\usepackage{amsfonts}
\usepackage{braket}
\usepackage{graphicx}
\usepackage[usenames,dvipsnames,svgnames,table]{xcolor}

\usepackage[colorlinks=true,linkcolor=blue, citecolor=blue, bookmarks]{hyperref}
\usepackage{tikz,pgfplots}

\newcommand{\rhotp}{\rho_t}
\newcommand{\rhop}{\rho}
\newcommand{\rhohp}{\rho_h}

\newcommand{\rhots}{\sigma_t}
\newcommand{\rhos}{\sigma}
\newcommand{\rhohs}{\sigma_h}

\newcommand{\limth}{{\textstyle \lim_{\rm th}}}

\newcommand{\nn}{\nonumber \\}
\newcommand{\ignore}[1]{{}}
\newcommand{\be}{\begin{equation}}
\newcommand{\ee}{\end{equation}}
\def\doi{http://dx.doi.org/}

\tikzset{
    stationary-density-legend/.pic = {
        \draw[violet,  thick] (0,1.1) -- (1,1.1);
		\draw  node[right] at (-1.8,1.1) {$q=0$};
		\draw[blue,  thick] (0,0.7) -- (1,0.7);
		\draw  node[right] at (-1.8,0.7) {$q=-0.25$};
		\draw[red,  thick] (0,0.3) -- (1,0.3);
		\draw  node[right] at (-1.8,0.3) {$q=0.5 e^{i \frac{6\pi}{7}}$};
		\draw[orange,  thick] (0,-0.1) -- (1,-0.1);
		\draw  node[right] at (-1.8,-0.1) {$q=e^{i\frac{6\pi}{5}}$};
    }
}

\tikzset{
    stationary-density-legend-2/.pic = {
        \draw[violet,  thick] (0,1.35) -- (1,1.35);
		\draw  node[right] at (-3.2,1.35) {$\{a_j\}=\{ -{1},0,0\}$};
		\draw[blue,  thick] (0,0.9) -- (1,0.9);
		\draw  node[right] at (-3.2,0.9) {$\{a_j\}=\{{1}, -{1},0\}$};
		\draw[red,  thick] (0,0.45) -- (1,0.45);
		\draw  node[right] at (-3.2,0.5) {$\{a_j\}=\{1,1,-1\}$};
		\draw[orange,  thick] (0,0) -- (1,0);
		\draw  node[right] at (-3.2,0) {$\{a_j\}=\{e^{i\frac{\pi}{5}},1,e^{i\frac{\pi}{3}}\}$};
    }
}

\tikzset{
	yyentropy-data-plot/.pic = {
		\begin{axis}[width=11cm,height=7.6cm,
			xmin=-3.14159265359,xmax=3.14159265359,
			ymin=0,ymax=0.125
			]
			\addplot[domain=-3.14159265359:3.14159265359,red,thick] table {\yyet};
			\addplot[domain=-3.14159265359:3.14159265359,dodgerblue,thick] table {\yyef};
			\addplot[domain=-3.14159265359:3.14159265359,lightseagreen,thick] table {\yyes};
		\end{axis}
	}
}

\tikzset{
    yyentropy-legend/.pic = {
		\draw[red, thick] (0,0.7) -- (1,0.7);
		\draw  node[right] at (-1,0.7) {$\nu=2$};
		\draw[dodgerblue, thick] (0,0.35) -- (1,0.35);
		\draw  node[right] at (-1,0.35) {$\nu=5$};
		\draw[lightseagreen, thick] (0,0) -- (1,0);
		\draw  node[right] at (-1,0) {$\nu=6$};
    }
}

\definecolor{lightseagreen}{RGB}{60,179,113}
\definecolor{dodgerblue}{RGB}{30,144,255}

\begin{document}
\title{Quantum Quench in the Infinitely Repulsive Hubbard Model: the Stationary State}
\author{Bruno Bertini}
\address{SISSA and INFN, via Bonomea 265, 34136, Trieste, Italy}
\author{Elena Tartaglia}
\address{SISSA and INFN, via Bonomea 265, 34136, Trieste, Italy}
\author{Pasquale Calabrese}
\address{SISSA and INFN, via Bonomea 265, 34136, Trieste, Italy}

\begin{abstract}
We use the Quench Action approach to study the non-equilibrium dynamics after a quantum quench in the Hubbard model in the limit of infinite interaction. 
We identify a variety of low-entangled initial states for which we can directly compute the overlaps with the 
Hamiltonian's eigenstates. For these initial states, we analytically find the rapidity distributions of the stationary state characterising the expectation values of all local observables. Some of the initial states considered are not reflection symmetric and lead to non-symmetric rapidity distributions. To study such cases, we have to introduce a generalised form for the reduced entropy which measures the entropy restricted to states with non-zero overlap. The initial states considered are of direct experimental realisability and also represent ideal candidates for studying non-equilibrium dynamics in the Hubbard model for finite interactions.   
\end{abstract}
\maketitle

\section{Introduction}

Integrable models constitute a fundamental milestone for the theoretical understanding of condensed matter physics. In these non-trivially interacting many-body systems the couplings are fine-tuned to restrict the scattering, giving rise to stable quasi-particle excitations. As a consequence, a large number of significant properties, \emph{e.g.}, the spectrum, can be calculated in an exact fashion. This valuable feature can be used to obtain a full characterisation of a vast range of fascinating many-body phenomena. The equilibrium properties of integrable models have been subject to a comprehensive investigation during the last fifty years, leading, \emph{e.g.}, to an exact description of their thermodynamics {\cite{takahashibook,korepinbook}}. In the course of this research, it has been realised that integrable models with non-trivial internal degrees of freedom, known as nested systems, have a much richer structure which severely complicates their treatment: they contain different species of quasi-particles carrying different physical information. This structure in turn allows one to observe a wider spectrum of physical phenomena, the most known example being the celebrated spin-charge separation~\cite{thebook}. Arguably, the most prominent member of  such  a family of systems is the one-dimensional fermionic Hubbard model~\cite{thebook}, which describes the lattice dynamics of interacting spinful fermions. The Hubbard model can be seen as the minimal model for describing real solids, taking into account many-body effects and going beyond the simple band theory.    

The interest in integrable models acquired renewed strength during the last decade, when tremendous experimental progress in cold atomic systems~\cite{bdz-08, ccgo-11, gbl-13, PolkonikovRMP11} triggered an intensive theoretical investigation of out-of-equilibrium dynamics in closed quantum many-body systems (see \emph{e.g.} the reviews~\cite{PolkonikovRMP11,cem-16, ef-16, CC:review, CaCh16, BD:review, caux_qareview, VR:review, impz-16, pret, VM:review, DeLucaArxiv16,GE15}). In this framework, integrable models played a key role. They have been successfully used for understanding the mechanisms of relaxation~\cite{ef-16, Caza06,BaSc08, CDEO08, scc-09, CaEF11, grd-10, mc-12, se-12,CoSC13, fe-13b, KoCC14,sc-14,KBC:Ising,spyros-15, ppv-17, dc-14, PC:anyons, PE:qbosons}, showing that local-in-space observables relax at large times. The statistical ensemble describing these stationary values is determined by the conserved operators of the system that constrain the dynamics. Exact studies carried out in integrable models~\cite{KSCCI:LL, POZS2-13, fe-13, fcec-14, POZS2-14, ga-14, IDWC15, iqc-16, PoVW17, pvc-16, PVCR17, emp-15, dlsb-15, doyon-15, bs-16, vecu-16, cardy-16, P:qbosons} revealed that such operators are the \emph{local} and \emph{quasi-local} conservation laws of the system~\cite{impz-16}; see also \cite{prosen-11, ip-12, prosen-14, zmp-16, imp-15, ppsa-14, pv-16}. In particular, stationary values of local observables are computed according to a standard Gibbs ensemble if the set of these charges is reduced to solely the Hamiltonian. Due to their restricted scattering, integrable models possess a macroscopic number of such conservation laws, and stationary values of local observables are generically non-thermal. These values are described by the so-called generalised Gibbs ensemble (GGE)~\cite{RigolPRL07} and bear much more information on the initial configuration compared to the standard thermal states. When a very weak integrability breaking perturbation is added to the system, a remarkable transition between GGE values and thermal values is observed in the time evolution of local observables. This phenomenon has been named ``prethermalization''~\cite{pret, MK:prethermalization,RoschPRL08,KollarPRB11,worm13,MarcuzziPRL13,EsslerPRB14,NIC14,fagotti-14,konik14,BF15,CTGM:pret,knap15,SmacchiaPRB15,BEGR:PRL,FC15,MenegozJStatMech15,mmgs-16,cgdm-17,BEGR:long,af-17,fnr-17,bbw-04}. Integrable models have also been used to study the entanglement dynamics~\cite{calabrese_evolution_2005, CC:review, alba_entanglement_2016, kauf, d-17, CA17} and more recently to obtain a full description of transport problems~\cite{BD:review, VM:review,CADY:hydro, BCDF:transport, BeFa16, F16, DoYo16, DoSY17, DoYC17, DDKY17, BVKM17-2, BVKM17, DeLucaAlvise, PDCBF:transport, A:entanglementtransport}. Most of the studies have been centred on integrable models without internal degrees of freedom, as for example the Lieb-Liniger gas or the XXZ spin chain. These models have been used to develop and test many different methods exploiting the powerful structure underlining integrability to find exact results~\cite{cc-06, fcc-09, cg-11, fm-10, pozsgay-11, BeSE14, ia-12,  mussardo-13, la-14, d-14, alba-2015, cubero-16,iqdb-16,caux_time_2013,kctc-17}. An example worth mentioning is the Quench Action approach~\cite{caux_time_2013}, which allows for a full determination of the large-time stationary state with no need for detailed knowledge of the full set of conserved charges~\cite{dwbc-14, BePC16, wdbf-14,QA:XXZ, PMWK14, mestyan_quenching_2015, ac-16qa, dc-14, pce-16, NaPC15, Bucc16, npg-17}. The price to pay is that it requires detailed microscopic information on the initial configuration: one has to know the overlaps between the initial state and the eigenstates of the Hamiltonian.

Until very recently~\cite{RCKPRL, RCK-16, kph-17,nestedquench, IDN:hubbard}, nested systems remained essentially neglected in out-of-equilibrium studies, mostly because of the technical difficulties inherent in their treatment. Investigating the out-of-equilibrium dynamics of nested systems is, however, a task of primary interest: it is very reasonable to think that nested systems will show new and interesting physics also in out-of-equilibrium situations, just as they do in equilibrium. Furthermore, we expect to see signatures of their equilibrium special features, as occurs for other peculiar systems \cite{kww-06,kctc-17,pce-16}. After having achieved good experimental control and theoretical understanding of the dynamics in non-nested integrable models, the field is now ready to begin a systematic investigation of nested systems. These systems are currently well within the reach of cold atomic experiments, \emph{e.g.} \cite{liao,Fallani:tunablespin,boll,pars,hil,gbl-13}, and some of the methods developed in the past, \emph{e.g.} the Quench Action approach, can also be applied in the nested case. 

Here we begin to undertake this task and consider the analytic characterisation by means of integrability of the out-of-equilibrium dynamics of the one-dimensional Hubbard model~\cite{thebook}, which so far, to the best of our knowledge, has been studied only by numerical or approximate means, as in \cite{MK:prethermalization,ekw-09,sf-10,qkns-14,ymwr-14,roh-15,yr-17,sjhb-17}. 
However, to deal with the intrinsic difficulty of the model, we focus on the limit of infinite interaction. This limit allows us to avoid many technical complications and concentrate on one of the essential properties of nested systems: the presence of multiple species of quasi-particles carrying different physical information. We consider a large class of low-entangled initial states that  are not invariant under one-site translations. The simplest example of a state in this class is that with a ``nested N\'eel" structure: the fermions occupy one lattice position every two and one fermion every two has spin down. We average these states over a number of one-site translations to make them translationally invariant and analytically compute their overlap with the eigenstates of the Hamiltonian. We use this result to find the exact thermodynamic description of the large-time stationary state by means of the Quench Action approach, introducing a generalised reduced entropy to account for non-parity symmetric initial states. Our study should be regarded as a first step towards the investigation of the fully interacting case: the initial states we consider are the natural candidates for studying quenches to the finite interaction Hubbard, and the use of the Quench Action approach makes the extension to the fully interacting case more direct.

The paper is laid out as follows. In Section \ref{sec:model} we introduce the model considered, briefly reviewing its Bethe ansatz solution and its thermodynamic description. A detailed account of our quench protocol and the class of initial states considered is given in Section~\ref{sec:quenchprotocol}. In Section~\ref{sec:stationarystate} we compute the root densities of the large-time stationary state using the Quench Action approach. Finally, in Section~\ref{sec:conclusions} we report our conclusions. A few technical points are reported in the appendices.  

\section{Hubbard Model with Infinite Interaction}
\label{sec:model}

We consider a system of interacting fermions on a one-dimensional lattice whose dynamics are described by the Hubbard model~\cite{thebook}. The Hamiltonian reads as 
\begin{equation}
\hat H = -t\sum_{x=1}^{L}\sum_{\alpha=\uparrow,\downarrow} 
\left(c_{x,\alpha}^\dagger c^{\phantom{\dagger}}_{x-1,\alpha} + c_{x-1,\alpha}^\dagger c^{\phantom{\dagger}}_{x,\alpha}\right )
+U \sum_{x=1}^L n_{x,\uparrow} n_{x,\downarrow}\qquad c_{0,\alpha}=c_{L,\alpha}.
\label{eq:hubbardham}
\end{equation}
Here the fermionic operators $c_{x,\alpha}$, $c_{x,\alpha}^\dagger$ satisfy anti-commutation relations
\begin{equation}
\{ c_{x,\alpha}, c_{y,\beta} \} = \{ c_{x,\alpha}^\dagger, c_{y,\beta}^\dagger \} = 0, \qquad
\{ c_{x,\alpha}, c_{y,\beta}^\dagger \} = \delta_{x,y}\delta_{\alpha,\beta}\,,
\end{equation}
and the number operators $n_{x,\alpha} = c_{x,\alpha}^\dagger c_{x,\alpha}$ count the number of particles with spin $\alpha$ at site $x$. The vacuum state is defined in the usual fashion as that annihilated by the fermionic operator $c_{x,\alpha}$
\begin{equation}
c_{x,\alpha} \ket{0} = 0, \qquad x=1,2,\ldots,L, \qquad \alpha = \uparrow, \downarrow.
\end{equation}
We also introduce the number operator $\hat N$ and the spin down number operator $\hat M$, defined as 
\be
\hat N=\sum_{x=1}^{L}\sum_{\alpha=\uparrow,\downarrow} c_{x,\alpha}^\dagger c^{\phantom{\dagger}}_{x,\alpha}\,,\qquad\qquad\hat M=\sum_{x=1}^{L}c_{x,\downarrow}^\dagger c^{\phantom{\dagger}}_{x,\downarrow}\,.
\ee
Any site on the lattice can have no particles, a fermion with spin up, a fermion with spin down or two fermions with different spin. These states at site $x$ are constructed by acting with the fermionic creation operators on the vacuum as follows
\begin{equation}
\ket{0}, \qquad c_{x,\uparrow}^\dagger\ket{0}, \qquad c_{x,\downarrow}^\dagger\ket{0}, 
\qquad c_{x,\uparrow}^\dagger c_{x,\downarrow}^\dagger\ket{0}\,,
\end{equation}
Since at every site there are $4$ possible states, the total Hilbert space has $4^L$ states. The states that have at least one site with two particles
are called \emph{double occupancy} states; note that these states are the only ones affected by the interaction term. 

In this paper, we are interested in the limit of infinite interaction $U\rightarrow\infty$. In this limit, the double occupancy states have infinite energy and therefore become unphysical. The physical Hilbert space is restricted to contain no double occupancy states and has dimension $3^L$. The effective Hamiltonian describing the dynamics in this limit is written as 
\begin{equation}
\hat H_\infty = -t\,\mathcal{P}\!\left[
\sum_{x=1}^L \sum_{\alpha=\uparrow, \downarrow} 
\left(c_{x,\alpha}^\dagger c^{\phantom{\dag}}_{x-1,\alpha} + c_{x-1,\alpha}^\dagger c^{\phantom{\dag}}_{x,\alpha}\right )
\right]\!\!\mathcal{P}, \qquad
\mathcal{P} = \prod_{x=1}^L (1-n_{x,\uparrow} n_{x,\downarrow}).
\label{eq:ham-inf-int}
\end{equation}
This Hamiltonian is known in the literature as the $t$\,-\,0 model~\cite{thebook} and is non-trivial only if the number of particles is less than $L$. In Ref.~\cite{GM:infUHubbard} it has been shown that the algebraic structure of \eqref{eq:ham-inf-int} is that of a graded $SU(3)$ generalisation of the XX spin chain. A mapping of \eqref{eq:ham-inf-int} to a non-interacting spinless fermionic Hamiltonian is presented in Ref.~\cite{K:Hubbardmapping} (see also Ref.~\cite{K:Operators}) in the case of open boundary conditions. This mapping is very useful to compute the time evolution of certain local observables and will be exploited in future works. Here instead we focus on the determination of the stationary properties after quantum quenches; we do this by adopting the Quench Action approach as it is directly generalisable to the fully interacting case. To this aim, we construct the eigenstates of \eqref{eq:ham-inf-int} via   coordinate Bethe ansatz and describe the thermodynamic limit using thermodynamic Bethe ansatz (TBA); these tasks are carried out in the following subsections. For convenience, in the remaining part of this paper we set $t=1$. 

\subsection{Eigenstates of the Hamiltonian}
\label{sec:eigenstates}
The eigenstates of the Hamiltonian can be found by \emph{nested} Bethe Ansatz~\cite{gaudinyang-67}. The solution was first presented in \cite{IPA:j0model} (see also \cite{thebook} for a detailed construction of the eigenstates of \eqref{eq:hubbardham}). Here we will briefly summarise the main steps. First we introduce the Bethe states
\begin{equation}
\ket{\Psi_{N,M}(\boldsymbol k;\boldsymbol \lambda)} = \sum_{1\leq z_1 < \ldots < z_N \leq L} \sum_{\alpha_1,\ldots,\alpha_N = \uparrow,
\downarrow}  \chi_{N,M}^{\boldsymbol \alpha}(\boldsymbol z|\boldsymbol k;\boldsymbol \lambda) 
\ c_{z_1,\alpha_1}^\dagger\ldots c_{z_N,\alpha_N}^\dagger\ket{0}\,,
\label{eq:estate-inf-int}
\end{equation}
where $\boldsymbol z=\{z_1, \ldots, z_N\}$, $\boldsymbol \alpha=\{\alpha_1,\ldots,\alpha_N\}$. The states $\ket{\Psi_{N,M}(\boldsymbol k;\boldsymbol \lambda)}$ live in the subspace with fixed particle number $N$ and spin down particle number $M$;  they are parametrised by the two sets of rapidities ${\boldsymbol k=\{k_i\}_{i=1,\ldots,N}}$ and ${\boldsymbol \lambda=\{\lambda_i\}_{i=1,\ldots,M}}$. The goal is to find appropriate wavefunctions $\chi_{N,M}^{\boldsymbol \alpha}(\boldsymbol z|\boldsymbol k;\boldsymbol \lambda)$ such that the Bethe states form a complete set of eigenstates of the Hamiltonian in the sector with fixed $N$ and $M$. Note that the Hamiltonian can be diagonalised in the sectors of fixed $N$ and $M$, since it conserves the number of particles and spin. This means that applying the Hamiltonian to a state of the form \eqref{eq:estate-inf-int} would give a state of the same form with the replacement
\be
\chi_{N,M}^{\boldsymbol \alpha}(\boldsymbol z|\boldsymbol k;\boldsymbol \lambda) \longmapsto
(H^{(L)}_{\infty,N}\chi_{N,M}^{\boldsymbol \alpha})(\boldsymbol z|\boldsymbol k;\boldsymbol \lambda)\,,
\ee
for some operator $H^{(L)}_{\infty,N}$, known as the first quantised Hamiltonian in a finite volume $L$. 

The problem of determining the wavefunction is split into two steps. First one requires $\chi_{N,M}^{\boldsymbol \alpha}(\boldsymbol z|\boldsymbol k;\boldsymbol \lambda)$ to be an eigenstate of $H^{(\infty)}_{\infty,N}$, the first quantised Hamiltonian on the infinite line, and then one imposes periodic boundary conditions on it 
\be
\chi_{N,M}^{\boldsymbol \alpha}(z_1,\dots,z_j+L,\ldots, z_N|\boldsymbol k;\boldsymbol \lambda)=\chi_{N,M}^{\boldsymbol \alpha}(z_1,\dots,z_j,\ldots, z_N|\boldsymbol k;\boldsymbol \lambda)\,,\qquad \forall\, j=1,\ldots,N\,.
\ee
In the case of infinite interaction, the first step yields the following wavefunction
\begin{equation}
\chi_{N,M}^{\alpha_1,\ldots,\alpha_N}(\boldsymbol z|\boldsymbol k;\boldsymbol \lambda) 
= \frac{1}{L^{N/2}} \left(\sum_{\mathcal P\in\mathcal S_N}
\xi_{N,M}^{\alpha_{\mathcal P(1)}, \ldots, \alpha_{\mathcal P(N)}}(\boldsymbol \lambda)\,\theta(z_{\mathcal P(1)}<\ldots<z_{\mathcal P(N)})\right) {\det}_N\{e^{i k_a z_b}\}\,.
\label{eq:wavefunction}
\end{equation}
Here $\mathcal{S}_N$ is the group of permutations of $N$ elements; $\theta(z_{1}<\ldots<z_{N})$ gives $1$ if $z_{1}<\ldots<z_{N}$ and $0$ otherwise; and  ${\det}_N\{e^{i k_a z_b}\}$ is the determinant of the $N\times N$ matrix with elements $e^{i k_a z_b}$ for $a,b = 1,\ldots,N$. Finally, $\xi_{N,M}^{\boldsymbol \alpha}(\boldsymbol \lambda)$ is an arbitrary coefficient: given a string of $N$ spins $\alpha_1...\alpha_N$, of which $M$ are down, it returns a number. Since $\xi_{N,M}^{\boldsymbol \alpha}(\boldsymbol \lambda)$ is arbitrary, any choice yielding a complete set of vectors $\ket{\Psi_{N,M}(\boldsymbol k;\boldsymbol \lambda)}$ is allowed. Here we follow Refs.~\cite{IPA:j0model, IP:twocomponentgases} and take as $\xi_{N,M}^{\boldsymbol \alpha}(\boldsymbol \lambda)$ the wavefunction of the periodic XX spin chain with $N$ spins and $M$ down spins. Namely, we choose 
\begin{equation}
\xi_{N,M}^{\boldsymbol \alpha}(\boldsymbol \lambda) = \frac{1}{N^{M/2}}{\det}_M\{e^{i \lambda_a n_b}\}\,,
\label{eq:choicexi}
\end{equation}
where $n_i$ gives the position of the $i^\text{th}$ down spin in $\{\alpha_1, \ldots, \alpha_N\}$ and the rapidities $\lambda_i$ satisfy
\be
e^{i \lambda_b N} = (-1)^{M+1}, \qquad\qquad b= 1,\ldots,M\,.
\label{eq:bethe1}
\ee
Imposing now periodic boundary conditions on $\chi_{N,M}^{\boldsymbol\alpha}(\boldsymbol z|\boldsymbol k;\boldsymbol \lambda)$ yields 
\be
e^{i k_a L} = e^{i \Lambda}, \quad\qquad\qquad\qquad a= 1,\ldots,N\,,
\label{eq:bethe2}
\ee
where
\be
\Lambda= \sum_{j=1}^{M}\lambda_j\,.
\ee
The states \eqref{eq:estate-inf-int} with $\chi_{N,M}^{\boldsymbol \alpha}(\boldsymbol z|\boldsymbol k;\boldsymbol \lambda)$ given in \eqref{eq:wavefunction} and $\xi_{N,M}^{\boldsymbol \alpha}(\boldsymbol \lambda)$ given in \eqref{eq:choicexi} are a set of eigenstates of the Hamiltonian $\hat H_\infty$ if the rapidities satisfy the quantisation conditions \eqref{eq:bethe1} and \eqref{eq:bethe2}. Note the normalisations have been chosen such that 
\be
\braket{\Psi_{N,M}(\boldsymbol k;\boldsymbol \lambda)|\Psi_{N',M'}(\boldsymbol k';\boldsymbol \lambda')}=\delta_{N,N'}\delta_{M,M'}\delta_{\boldsymbol k, \boldsymbol k'}\delta_{\boldsymbol \lambda,\boldsymbol \lambda'}\,.
\ee
The quantisation conditions \eqref{eq:bethe1} and \eqref{eq:bethe2} are known as Bethe equations, and in logarithmic form they read as 
\begin{subequations}
\begin{align}
x(k_a) &= \frac{2\pi}{L}I_a\,, & I_a&\in\mathbb{Z}\cap[-\tfrac{L}{2},\tfrac{L}{2}), & a&=1,\ldots,N,\label{eq:quantcondk}\\
y(\lambda_b) &= \frac{2\pi}{L}J_b, & J_b&\in
\begin{cases}
\mathbb{Z}\cap[-\tfrac{N}{2},\tfrac{N}{2})\,\,\quad\qquad M\quad\text{odd}\\
\mathbb{Z}_{1/2}\cap[-\tfrac{N}{2},\tfrac{N}{2})\qquad M\quad\text{even}
\end{cases} 
 & b&=1,\ldots,M\,.
\end{align}
\label{eq:rapidities}
\end{subequations}
Here $\mathbb Z_{1/2}$ is the set of half odd integers, and we introduced the counting functions 
\begin{align}
x(k) &=  k -\frac{\Lambda \!\!\!\mod 2\pi}{L},\qquad\qquad\qquad
y(\lambda) =\frac{L}{N} \lambda.
\label{eq:counting-functions}
\end{align}
The eigenvalues of the Hamiltonian \eqref{eq:ham-inf-int} on eigenstates \eqref{eq:estate-inf-int} are given by 
\begin{equation}
E=\sum_{a=1}^N \varepsilon(k_a), \qquad \varepsilon(k) = -2\cos(k).
\end{equation}
It is also useful to define the translation operator $\hat T$ and the reflection operator $\hat R$ such that 
\be
\hat Tc^\dagger_{x,\alpha} \hat T^\dagger=c^\dagger_{x+1,\alpha}\,,\qquad\qquad\qquad\hat Rc^\dagger_{x,\alpha} \hat R^\dagger=c^\dagger_{L+1-x,\alpha}\,.
\label{eq:translation-op}
\ee
Applying $\hat T$ and $\hat R$ to the eigenstates \eqref{eq:estate-inf-int} gives
\begin{equation}
\hat T\ket{\Psi_{N,M}({\boldsymbol k};\mathbf{\boldsymbol\lambda})} = e^{-iK} \ket{\Psi_{N,M}(\boldsymbol k;\boldsymbol \lambda)},\qquad \hat R\ket{\Psi_{N,M}(\boldsymbol k;\boldsymbol\lambda)} = (-1)^{\lfloor M/2\rfloor}e^{i(\Lambda+K)}\ket{\Psi_{N,M}(-\boldsymbol k;-\boldsymbol\lambda)}\,,
\end{equation}
where 
\be
K=\left[\sum_{a=1}^{N} k_a\right]\!\text{mod}\,2\pi\,
\ee
is the momentum of the states \eqref{eq:estate-inf-int}.

\subsection{Thermodynamic Description}

Our aim is to study the system in the thermodynamic limit 
\begin{equation}
L,\,N,\,M \rightarrow \infty, \qquad\qquad n\equiv\frac{N}{L}, \quad m\equiv\frac{M}{L} \text{ fixed}\,,
\end{equation}
where $N$ is the number of particles and $M$ is the number of spin down particles. In the following, we denote this limit by $\limth$. In the thermodynamic limit, both species of rapidities become dense in the interval $[-\pi,\pi)$, and the description is conveniently carried out in terms of their densities, called root densities. The root densities describing the state $\ket{\Psi_{N,M}(\boldsymbol k;\boldsymbol \lambda)}$ in the thermodynamic limit are defined by 
\be
\rhop(k_j) \equiv \limth \frac{1}{L(k_{j+1}-k_j)}\,, \qquad\qquad
\rhos(\lambda_j) \equiv \limth \frac{1}{L(\lambda_{j+1}-\lambda_j)}\,.
\ee
For the thermodynamic description it is also convenient to consider ``holes" $\{k^h_i\},\{\lambda^h_j\}$. Holes correspond to values of the rapidities which are in principle allowed by the quantisation conditions but for which there is no particle in the state. They also become dense in the thermodynamic limit, and their densities $\rhohp(k)$ and $\rhohs(k)$ are defined in an analogous fashion. Finally, we define the total root densities of the two species of rapidities as
\be
\rhotp(k)=\rhop(k)+\rhohp(k)\,,\qquad\qquad\rhots(\lambda)=\rhos(\lambda)+\rhohs(\lambda)\,.
\ee
Using the definition of the counting functions $x(k)$ and $y(\lambda)$ we find 
\begin{equation}
\rhotp(k) = \frac{1}{2\pi}x'(k)  = \frac{1}{2\pi}, \qquad \rhots(\lambda) =\frac{1}{2 \pi} y'(\lambda)= \frac{1}{2\pi}\int_{-\pi}^\pi\!\!\!{\rm d}k\,\rhop(k)\,.
\label{eq:total-rapidity-densities}
\end{equation}
These equations, known as TBA equations, connect the root densities and densities of holes. Note that for the case under examination the TBA equations are almost those of a free theory---the equation for $\rhotp(k)$ is independent of $\rhop(k)$ and $\rhos(\lambda)$, while that of $\rhots(\lambda)$ depends on $\rhop(k)$ only through the density $n=\int_{-\pi}^\pi\!{\rm d}k\,\rhop(k)$.

The basic assumption of this thermodynamic description is that the root densities fully characterise a macrostate of the system in the thermodynamic limit, \emph{i.e.} they characterise a set of many different eigenstates of the Hamiltonian that have the same expectation values for all the local operators in the thermodynamic limit. 
In our case for example, the densities of energy, particles and spin down particles of a macrostate are written in terms of the root densities as
\begin{align}
e &= \limth\frac{E}{L} = \int_{-\pi}^\pi \ \mathrm{d}k \ \varepsilon(k)\rhop(k) = -2\int_{-\pi}^\pi \ \mathrm{d}k \ \cos(k)\rhop(k)\,,\\
n &= \limth \frac{N}{L} = \int_{-\pi}^\pi \ \mathrm{d}k \ \rhop(k)\,,
\qquad m =\limth \frac{M}{L} = \int_{-\pi}^\pi \!\! \mathrm{d}k \ \rhos(k)\,.
\end{align}
The number $N_{\boldsymbol \rho}$ of eigenstates of the Hamiltonian corresponding to a particular macrostate is exponentially large in~$L$
\be
N_{\boldsymbol \rho}\sim e^{L s_{\text{YY}}[\boldsymbol \rho]}\,,
\ee
where the functional $s_{\text{YY}}[\boldsymbol \rho]$ is known as Yang-Yang entropy density. In our case we have  
\begin{align}
s_\text{YY}[\boldsymbol \rhop] = s_\text{YY}[ \rhop, \rhos]  \equiv \int_{-\pi}^\pi \mathrm{d}k 
\left\{ s_\text{YY}[\rhop,\rhohp](k)+ s_\text{YY}[\rhos,\rhohs](k)\right\}\,,
\label{eq:yang-yang-entropy}
\end{align}
where 
\be
s_\text{YY}[f,g](k)\equiv(f(k)+g(k))\log(f(k)+g(k)) - f(k)\log f(k) - g(k)\log g(k)\,.
\label{eq:yangyang}
\ee

\section{Quench Protocol}
\label{sec:quenchprotocol}

Our strategy is to generate out-of-equilibrium dynamics using a standard quantum quench. We take the system in an initial state $\ket{\Psi_0}$, which can be thought of as the ground state of a translationally invariant Hamiltonian, and evolve it by means of the Hamiltonian $\hat H_\infty$. The focus here is on determining the stationary state $\ket{\Psi^{\infty}_s}$ that describes expectation values of local observables in the thermodynamic limit at infinite times. An interesting question is whether this state corresponds to the infinite $U$ limit of the stationary state $\ket{\Psi^{U}_s}$, reached when evolving with the Hubbard Hamiltonian $\hat H$ (\emph{cf}.~\eqref{eq:hubbardham}). This happens if the limit of infinite time and that of infinite interaction commute, namely
\begin{equation}
\lim_{U\rightarrow\infty}\lim_{t\rightarrow\infty} \frac{\braket{\Psi_0|e^{i\hat Ht}\mathcal{O}e^{-i \hat H t}|\Psi_0}}{\braket{\Psi_0|\Psi_0}}= \lim_{t\rightarrow\infty}\lim_{U\rightarrow\infty} \frac{\braket{\Psi_0|e^{i\hat Ht}\mathcal{O}e^{-i \hat H t}|\Psi_0}}{\braket{\Psi_0|\Psi_0}}\,,
\label{eq:assumption}
\end{equation}
for any local observable~$\mathcal O$. This has been shown to happen in the strong coupling limit of the Lieb-Liniger model both in the bosonic~\cite{KSCCI:LL, KoCC14,dwbc-14,dwbc-14} and in the anyonic~\cite{PC:anyons} case.  One might presume the property \eqref{eq:assumption} to hold fairly generally for strong coupling limits, as they share many common features; \emph{e.g.} a part of the Hilbert space becomes unphysical and is projected away. In our case, however, the situation seems to be more complicated due to the nested structure of the system. The spin part of the Bethe states is indeed very different in the infinite $U$ case compared to the finite $U$ one: in the $U=\infty$ case it is arbitrary, while it is completely fixed for any finite interaction. When $U\rightarrow\infty$ the basis of the spin sector turns out to be given by that of the XXX spin-1/2 chain~\cite{os:strongcoupling}. Even if \eqref{eq:assumption} where not to hold for generic observables, our assumption is that $\ket{\Psi^{\infty}_s}$ would still describe a prerelaxation plateau~\cite{fagotti-14, BF15, FC15} appearing for $1\ll t\ll U$ in the time evolution of local observables under $\hat H$.  

To determine the stationary state root densities we adopt the Quench Action approach~\cite{caux_time_2013}. Before reviewing it, however, we introduce the family of initial states considered in this work.

\subsection{Initial States}
Let us start by introducing the state
\begin{equation}
 \ket{N}=\prod_{j=1}^{L/4}c^\dagger_{4 j-2,\uparrow}c^\dagger_{4j,\downarrow}\ket{0} = \bigotimes_{j=1}^{L/4}  \ket{\circ\uparrow\circ\downarrow}\,.
\label{eq:nestedneel}
\end{equation}
In the pictorial representation adopted here the $\circ$ denotes an empty site, the $\uparrow$ a particle with up spin and the $\downarrow$ a particle with down spin. The state $\ket{N}$ is well defined when the chain length $L$ is multiple of 4, and it lives in the sector of fixed particle density $n=N/L=1/2$ and spin density $m=M/L=1/4$. We see that this state can be thought of as a  ``nested N\'eel state", \emph{i.e.}, there is a N\'eel-like structure in both particle occupation and spin---there is a particle every two sites and every two particles there is one with spin down. 

The state \eqref{eq:nestedneel} can be generalised by varying the particle and spin densities.  Consider a state with the same form of $\ket{N}$ but with a particle every $\nu$ sites and a spin down every $\nu\mu$ sites, where $\nu,\mu\geq2$ are integers. In the pictorial representation the state reads as
\begin{equation}
\ket{N_{\nu\mu}} =  \bigotimes_{j=1}^{L/(\mu\nu)}  \ket{\overbrace{\underbrace{\circ\cdots\circ\uparrow}_\nu\cdots \underbrace{\circ\cdots\circ\uparrow}_\nu \underbrace{\circ\cdots\circ\downarrow}_\nu}^{\nu\mu}}\,. 
\end{equation}
A further generalisation can be obtained by allowing the particles and down spins to occur in any position of the regions indicated by the braces instead of being restricted to occur in the last positions. Let us construct these states by first placing the particles and then fixing their spin. Each region of $\nu$ positions indicated by the lower braces can contain exactly one particle, which can be placed at any of the $\nu$ positions (instead of just in the last position as above). A particle can occur at each position $m=1,2,...,\nu$ with amplitude $a_{\nu-m}$. Having now placed $\mu$ particles in the region of $\nu\mu$ positions indicated by the upper brace, we can fix any one of them to be spin down and the rest to be spin up (instead of only the last being spin down as above). The $k$th particle can have spin down with amplitude $b_{\mu-k}$, for $k=1,2,..,\mu$. The state is explicitly written as  
\begin{align}
\ket{\Phi^{\nu\mu}_{\{a_1,\ldots,a_{\nu-1}\} \{b_1,\ldots,b_{\mu-1}\}}} &=\prod_{j=0}^{L/(\nu\mu)-1}\left[
\sum_{k=0}^{\mu-1}
 \frac{b_{k}}{\sqrt{\sum_{\ell=0}^{\mu-1}|b_\ell|^2}} 
\prod_{\ell=1}^\mu \left(\sum_{m=0}^{\nu-1}  \frac{a_{m}}{\sqrt{\sum_{\ell=0}^{\nu-1}|a_\ell|^2}}  c^\dagger_{\nu\mu j+\ell\nu-m,s(\delta_{k,\mu-\ell})}
\right)\right]\ket{0}\,.
\label{eq:initial2}
\end{align}
Here, for convenience, we introduced $s(x)$ such that $s(0)=\,\uparrow$ and $s(1)=\,\downarrow$. These states are well defined when the chain length $L$ is multiple of $\nu\mu$. Note that, since the states are normalised, one of the $a_m$ and one of the $b_k$ are redundant, \emph{i.e.}, they are not necessary to specify the state up to a global phase. We then set $a_0=1$, $b_0=1$ and specify the state by means of the sets of complex parameters $\{a_m\}$ and $\{b_k\}$ with $m=1,\ldots,\nu-1$ and $k=1,\ldots,\mu-1$. The special cases discussed above are recovered from \eqref{eq:initial2} as follows 
\be
\ket{N} = \ket{\Phi^{22}_{00}}\qquad\qquad\qquad\ket{N_{\nu\mu}} = \ket{\Phi^{\nu\mu}_{\underbrace{\mbox{\small\{0\ldots0\}}}_{\nu-1}
\underbrace{\mbox{\small\{0\ldots0\}}}_{\mu-1}}} \,.
\ee
Another interesting special case of \eqref{eq:initial2} is 
\begin{align}
\ket{\Phi^{22}_{qp}} &= \bigotimes_{j=1}^{L/4}\frac{1}{\sqrt{1+|p|^{2}}}
\left[
\frac{\ket{\circ \uparrow}+q\ket{\uparrow\circ}}{\sqrt{1+|q|^{2}}}\otimes\frac{\ket{\circ \downarrow}+q\ket{\downarrow\circ}}{\sqrt{1+|q|^{2}}}
+p
\left(\frac{\ket{\circ \downarrow}+q\ket{\downarrow\circ}}{\sqrt{1+|q|^{2}}}\otimes
\frac{\ket{\circ \uparrow}+q\ket{\uparrow\circ}}{\sqrt{1+|q|^{2}}}\right)
\right]\,.
\label{eq:initial1}
\end{align} 
This state has a $q$-dimer structure in the particle occupation and a $p$-dimer structure in the spin.

As initial states we take the translationally invariant version of the states \eqref{eq:initial2}, constructed by acting with the translation operator $\hat T$ (\emph{cf}. \eqref{eq:translation-op}) as follows 
\begin{equation}
\ket{\Psi^{\nu\mu}_{\{a_m\}\{b_k\}}} = \frac{1}{\sqrt{\nu\mu}}\left(\sum_{j=0}^{\nu\mu-1} \hat T^j\right)\!\ket{\Phi^{\nu\mu}_{\{a_m\}\{b_k\}}}\,, \qquad \hat T^0 = I\,.
\label{eq:initial2-ti}
\end{equation}
We restrict to states such that $\hat T^{\nu\mu}$ acts as the identity. This occurs when $\mu$ is even or the number of down spins $L/(\nu\mu)$ is odd.  We choose to consider translationally invariant initial states to ensure the applicability of the TBA formalism. It is indeed widely believed that all the macrostates contributing to the time evolution of a translationally invariant state can be described in terms of the sole root densities: no exceptions are known. The symmetrisation, however, causes the states \eqref{eq:initial2-ti} to lose the cluster decomposition properties possessed by the states \eqref{eq:initial2}. Nevertheless, if translational symmetry is restored for the states \eqref{eq:initial2}, this implies that the cluster decomposition property is restored for the states \eqref{eq:initial2-ti}. 
This happens, \emph{e.g.}, in the XX model when evolving from the analogues of the states \eqref{eq:initial1} (spinless fermions and $p=0$)~\cite{FC15}. 

Note that for generic values of the parameters $\{a_m\}$ and $\{b_k\}$ the states \eqref{eq:initial2-ti} are not reflection symmetric. Specifically, we have 
\be
\hat R \ket{\Psi^{\nu\mu}_{
\{a_1,\ldots,a_{\nu-1}\}
\{b_1,\ldots,b_{\mu-1}\}}}=
(-1)^{\lfloor L/(2\nu)\rfloor} 
\ket{\Psi^{\nu\mu}_{
\left\{ \tfrac{a_{\nu-2}}{a_{\nu-1}},\ldots, \tfrac{a_{1}}{a_{\nu-1}}, 
\tfrac{1}{a_{\nu-1}} \right\}
\left\{ \tfrac{b_{\mu-2}}{b_{\mu-1}},\ldots, \tfrac{b_{1}}{b_{\mu-1}}, 
\tfrac{1}{b_{\mu-1}} \right\}
 }}.
\ee
In particular $\ket{\Psi^{22}_{qp}}$ is an eigenstate of the reflection operator only for $p=0,1$ and $q=0,1$. 

By construction, the states \eqref{eq:initial2-ti} live in the sector of fixed particle and down spin particle densities
\be
n=\frac{N}{L} = \frac{1}{\nu}\,,\qquad\qquad m=\frac{M}{L} = \frac{1}{\nu\mu}\,.
\label{eq:ISdensities}
\ee 
Moreover, using that there are no double occupancies in the states \eqref{eq:initial2-ti} we can compute the energy density of the state by getting rid of the projectors in \eqref{eq:ham-inf-int}. After a simple but tedious calculation we find 
\be
e= \frac{1}{L}\braket{\Psi^{\nu\mu}_{\{a_m\}\{b_k\}}|\hat H_\infty |\Psi^{\nu\mu}_{\{a_m\}\{b_k\}}}=-2 \sum_{\alpha=\uparrow, \downarrow} \text{Re}\left[\braket{\Psi^{\nu\mu}_{\{a_m\}\{b_k\}}|c_{2,\alpha}^\dagger c^{\phantom{\dag}}_{1,\alpha} |\Psi^{\nu\mu}_{\{a_m\}\{b_k\}}}\right]= -\frac{2}{\nu}\frac{\sum_{n=1}^{\nu-1} \text{Re}\left[{a^*_n  a_{n-1}}\right] }{{\sum_{\ell=0}^{\nu-1}|a_\ell|^2}}\,. 
\label{eq:ISenergydensity}
\ee

\subsection{Quench Action Approach}
\label{sec:QAA}

The main idea underlying the Quench Action approach is that, in the thermodynamic limit, one can characterise the time evolution of local observables in integrable models in terms of a single eigenstate of the Hamiltonian called the ``representative state". The representative state, or more precisely the corresponding root densities, are generically determined by a set of integral equations. This is very much like what happens in the TBA treatment of the thermal equilibrium state, where, in the thermodynamic limit, the density matrix of the canonical ensemble is represented in terms of a single eigenstate of the Hamiltonian, whose root densities are characterised by a set of integral equations. A pedagogical introduction to the Quench Action approach is given in the recent review~\cite{caux_qareview}. Here we are going to briefly sketch the main aspects and fix the relevant notation.

Let us consider the time evolution of the expectation value of a generic local observable $\mathcal O$ according to a generic integrable Hamiltonian $H$
\begin{equation}
\braket{\mathcal{O}(t)} \equiv \frac{\braket{\Psi_0|\mathcal{O}(t)|\Psi_0}}{\braket{\Psi_0|\Psi_0}},
\label{eq:expect_loc_obs}
\end{equation}
where $\mathcal O(t) = e^{iH t} \mathcal O e^{-iH t}$ and $\ket{\Psi_0}$ is the initial state. It was argued in Ref.~\cite{caux_time_2013} that in the thermodynamic limit 
\begin{equation}
\limth \braket{\mathcal{O}(t)}
= \limth \left[ \frac{\braket{\Psi_0|\mathcal{O}(t)|\Psi_s}}{2\braket{\Psi_0|\Psi_s}} + \frac{\braket{\Psi_s|\mathcal{O}(t)|\Psi_0}}{2\braket{\Psi_s|\Psi_0}}\right]\,,
\label{eq:repstate1}
\end{equation}
where $\ket{\Psi_s}$ is a particular eigenstate of the Hamiltonian known as the representative state. The relation \eqref{eq:repstate1} holds for any time $t>0$, and by taking the infinite time limit one finds that $\ket{\Psi_s}$ gives a representation of the stationary state describing expectation values of local observables at late times
\begin{equation}
\lim_{t\rightarrow\infty}\limth \braket{\mathcal{O}(t)} =
\limth \frac{\braket{\Psi_s|\mathcal{O}|\Psi_s}}{\braket{\Psi_s|\Psi_s}}\,.
\end{equation}
The representative eigenstate can be determined in the thermodynamic limit, when the eigenstates of the Hamiltonian are characterised by a set of root densities $\boldsymbol \rho$. The root densities $\boldsymbol \rho_s$ characterising the representative eigenstate are found by solving the following set of equations
\be
\frac{\delta\mathcal{F[\boldsymbol \rho]}}{\delta \boldsymbol \rho}\biggl|_{\boldsymbol \rho = \boldsymbol \rho_s} = 0\,,
\label{eq:saddle_point}
\ee
known as the saddle point equations. Here we introduced the Quench Action functional $\mathcal F[\boldsymbol \rho]$, which is defined as 
\be
\mathcal F[\boldsymbol \rho]=2 \mathcal{E}[\boldsymbol \rho]-s^\text{red}[\boldsymbol \rho]\,.
\ee
 The functional $\mathcal{E}[\boldsymbol \rho]$ appearing here is given by 
\begin{equation}
\mathcal{E}[\boldsymbol \rho] = - \limth \frac{1}{L}\log|\braket{\Phi_n|\Psi_{0}}|\,, 
\label{eq:initial-overlap}
\end{equation}
where $\ket{\Phi_n}$ is an eigenstate of $H$ corresponding in the limit to the set of root densities $\boldsymbol{\rho}$. The functional $s^{\text{red}}[\boldsymbol \rho]$ instead is the reduced entropy density, \emph{i.e.} the Yang-Yang entropy density (\emph{cf}. \eqref{eq:yangyang}) reduced to the states having non-zero overlap with the initial state $\ket{\Psi_0}$. In other words, for large but finite $L$ the exponential of the reduced entropy $L s^\text{red}[\boldsymbol \rho]$ gives the number of eigenstates of $H$ described by the same set of root densities $\boldsymbol \rho$ and having non-zero overlap with the initial state. Most of the studies carried out up to now~\cite{CA17, dwbc-14, BePC16, wdbf-14,QA:XXZ, PMWK14, mestyan_quenching_2015, ac-16qa, nestedquench, pce-16, NaPC15, Bucc16, npg-17} focused on initial states selecting root distributions that are symmetric around one point. In this case the reduced entropy turns out to be simply half of the Yang-Yang entropy \eqref{eq:yang-yang-entropy}. For the states \eqref{eq:initial2} this is generically not the case, and, accordingly, our reduced entropy $s_\text{red}[\boldsymbol \rhop]$ will not generically be a simple fraction of $s_\text{YY}[\boldsymbol \rhop]$ (\emph{cf}. Sec.~\ref{sec:reducedentropygen}).



\section{Determination of The Post-Quench Stationary State}
\label{sec:stationarystate}

Our goal here is to determine the set of root densities $\{\rhop_s(k),\rhos_s(\lambda)\}$ characterising the representative state. In order to do this, we write the explicit form of the Quench Action functional $\mathcal{F}[\rhop,\rhos]$ and solve the Equation~\eqref{eq:saddle_point}. As will be clear in the following, the construction of $\mathcal F [\rhop,\rhos]$ depends on the form of the initial state, because states in the class \eqref{eq:initial2-ti} with different  $\nu$ and $\mu$ give different reduced entropies. For the sake of clarity we start by considering the simpler subclass $\ket{\Psi^{22}_{qp}}$ and move to considering the general case later.   

\subsection{Initial states as $q$-dimers}

The construction of the Quench Action functional can be split in three main steps. First one has to find the overlaps between the initial states $\ket{\Psi^{22}_{qp}}$ and the Bethe states. Second, in the thermodynamic limit, one has to use the overlaps to construct $\mathcal{E}[\boldsymbol \rho]$. Finally, one constructs the reduced entropy $s^{\text{red}}[\boldsymbol \rho]$. We will undertake this task in the following subsections and present the solution to the saddle point equations in Sec.~\ref{sec:solutionqdimer}.  

\subsubsection{Overlaps with the Bethe States}
The overlaps between the states $\ket{\Psi^{22}_{qp}}$ and the eigenstates of the Hamiltonian are computed by using the simple form \eqref{eq:wavefunction}--\eqref{eq:choicexi} of Bethe states' wavefunctions $\ket{\Psi_{N,M}(\boldsymbol k;\boldsymbol \lambda)}$. After a simple calculation we find
\begin{equation}
\braket{\Psi^{22}_{qp}|\Psi_{N,M}(\boldsymbol k;\boldsymbol \lambda)}  = \frac{1}{2} \left(1+e^{iK}+e^{2iK}+e^{3iK}\right)\! \braket{\Phi^{22}_{qp}|\Psi_{N,M}(\boldsymbol k;\boldsymbol \lambda)}\,,\qquad K=\sum_{a=1}^{N} k_a\,,
\label{eq:initial1-overlap}
\end{equation}
where the overlap between $\ket{\Phi^{22}_{qp}}$ and the states $\ket{\Psi_{N,M}(\boldsymbol k;\boldsymbol \lambda)}$ is given by   
\begin{align}
\braket{\Phi^{22}_{qp}|\Psi_{N,M}(\boldsymbol k;\boldsymbol \lambda)}
&= 
\frac{\delta_{N,{L}/2}\delta_{M,{L}/4}}{L^{N/2} N^{M/2}}
\prod_{j=1}^N\!\left[\frac{1+q e^{-ik_j}}{\sqrt{1+|q|^2}}\right] \prod_{j=1}^{M}\!\left[\frac{1+p e^{-i\lambda_j}}{\sqrt{1+|p|^2}}\right] e^{i \Lambda (N+1)/2} {\det}_{N}\{e^{2 i \tilde k_ab} \} {\det}_{M}\{e^{2 i \lambda_ab} \}.
\label{eq:initial1-overlapphi}
\end{align}
Here we introduced 
\begin{equation}
\tilde{k}_a \equiv k_a - \frac{\Lambda\!\!\!\mod 2\pi}{L} = \frac{2\pi}{L}I_a\,, \qquad I_a\in\mathbb{Z}\cap[-\tfrac{L}{2},\tfrac{L}{2}-1], \qquad a=1,\ldots,N\,.
\label{eq:ktilde}
\end{equation}
Note that for generic $q$ and $p$ the overlaps can be easily related to the ones of the case $p=q=0$ analogously to what was done in Ref.~\cite{mestyan_quenching_2015} for the overlaps between the $q$-dimer states and the eigenstates of the XXZ spin chain. These overlaps have a very simple form: the Kronecker deltas fix $N=L/2$ and $M=L/4$ and the determinants in Eq.~\eqref{eq:initial1-overlapphi} are of the same form as those appearing in the overlaps between the N\'eel state and the eigenstates of the XX model. Therefore they can be treated as in Ref.~\cite{mazza_overlap_2016}. First one notes that the determinants are non-zero only when no two $\tilde{k}_a$ or $\lambda_a$ differ by $\pi$, to avoid two columns in the matrices being equal. Moreover, when the determinants are non-zero, their absolute value does not depend on the rapidities. This is because for all the cases where the determinants are non-zero the matrices are permutations of the same columns. Combining this with the condition that the term in brackets on the r.h.s. of \eqref{eq:initial1-overlap} must be non-zero for a non-zero overlap, we find
\be
\left|\frac{1+e^{iK}+e^{2iK}+e^{3iK}}{2L^{L/4} (L/2)^{L/8}}\,{\det}_{\frac{L}{2}}\{e^{2 i \tilde k_ab} \} {\det}_{\frac{L}{4}}\{e^{2 i \lambda_ab} \}\right|=
\begin{cases}
C\neq0, &\text{no two $\tilde{k}_a$ or $\lambda_a$ differ by $\pi$ $\wedge$ $K=0\,\text{mod}\,2\pi$}\,,\\
0, &\text{otherwise}\,.
\end{cases}
\label{eq:condition}
\ee 
This constant can be determined through a counting argument, rather than explicitly evaluating the determinants. Firstly, we note that equation \eqref{eq:condition} is nothing but the modulus of the overlap of the N\'{e}el state with the Bethe states $|\braket{\Psi^{22}_{00}|\Psi_{\frac{L}{2},\frac{L}{4}}}|$. Using the fact that the N\'{e}el state is normalised, we relate the constant $C$ to the number of Bethe states allowed by the condition \eqref{eq:condition} as follows 
\begin{align}
1 &= \braket{\Psi_{00}^{22}|\Psi_{00}^{22}} = \sum_{\boldsymbol k, \boldsymbol \lambda} |\braket{\Psi^{22}_{00}|\Psi_{N,M}(\boldsymbol k;\boldsymbol \lambda)}|^2
= (\# \text{ allowed Bethe states})\, C^2.
\end{align}
Determining $C$, therefore, reduces to counting the number of Bethe states which have non-zero overlap with the N\'{e}el state. 

We count the number of allowed Bethe states by first constructing states which satisfy the condition that no two rapidities of each type differ by $\pi$. We start by constructing one such state and then modifying it to construct the rest. The simplest one has all the allowed values of $\tilde{k}_a, \lambda_b\in[0,\pi)$, of which there are $L/2$ for the $\tilde{k}_a$ and $L/4$ for the $\lambda_b$. Then each rapidity can be shifted by subtracting $\pi$, giving a total of $2^{L/2}\times 2^{L/4}$ options for sets of rapidities whereby no two differ by $\pi$.
Now we must only take from these the states which satisfy $K=0 \!\!\mod2\pi$. Loosely, if we take one state with $K=0\!\!\mod2\pi$, all other states with the same value of $K$ modulo $2\pi$ are obtained by shifting an even number of particle rapidities by $\pi$ and an even number of spin rapidities by $\pi$. This means that a quarter of the $2^{L/2}\times 2^{L/4}$ states have $K=0\!\!\mod2\pi$, and the total number of allowed Bethe states is $2^{3L/4-2}$. A more rigorous argument is given in Appendix~\ref{app:counting-bethe-states}. From this we determine the constant $C$ to be
\begin{equation}
C = \frac{1}{\sqrt{\# \text{ allowed Bethe states }}} =  2^{-3L/8+1}\,.
\end{equation}
Note that the condition \eqref{eq:condition} imposes a macroscopic number of constraints on the distribution of the rapidities and becomes a constraint on the root densities in the thermodynamic limit. Only the states with root densities obeying this constraint will have non-zero overlap with the initial states $\ket{\Psi^{22}_{qp}}$. 

\subsubsection{Thermodynamic Limit of the Overlaps}
Let us consider the logarithm of the absolute value of the overlap \eqref{eq:initial1-overlap}, focusing on distributions of rapidities for which the overlap is non-zero. Using \eqref{eq:condition} we find 
\begin{align}
\log |\braket{\Psi^{22}_{qp}|\Psi_{\frac{L}{2},\frac{L}{4}}(\boldsymbol k;\boldsymbol \lambda)}|=& \frac{1}{2}\sum_{j=1}^{L/2}\log\left(1+\frac{2|q|\cos( k_j+\alpha)}{1+|q|^2}\right) +
\frac{1}{2}\sum_{j=1}^{L/4}\log\left(1+\frac{2 |p|\cos(\lambda_j+\beta)}{1+|p|^2}\right)\notag\\
&-\left(\frac{3L}{8}-1\right)\log2\,,
\end{align}
where we have expressed the terms involving $q$ and $p$ from \eqref{eq:initial1-overlapphi} in terms of trigonometric functions rather than complex exponentials, and we defined
\be
\alpha=-\text{arg}[q]\qquad\qquad\qquad\beta=-\text{arg}[p]\,.
\ee
In the thermodynamic limit we find   
\begin{align}
\mathcal{E}[ \rhop,\rhos] &= - \lim_{L\rightarrow\infty}\frac{1}{L}\log |\braket{\Psi^{22}_{qp}|\Psi_{\frac{L}{2},\frac{L}{4}}(\boldsymbol k;\boldsymbol \lambda)}|\notag\\
&=-\frac{1}{2}\int_{-\pi}^\pi \mathrm{d}k\left\{ \log\left(1+\frac{2|q|\cos(k+\alpha)}{1+|q|^2}\right)\rhop(k) +\log\left(1+\frac{2 |p|\cos(k+\beta)}{1+|p|^2}\right)\rhos(k) \right\}+\frac{3\log2}{8}.
\end{align}
In order for the overlap to be non-zero the root densities appearing here must satisfy the following relations 
\begin{equation}
\rhop(k) + \rhop(k-\pi) = \frac{1}{2\pi}, \qquad
\rhos(\lambda) + \rhos(\lambda-\pi) = \frac{1}{4\pi}, \qquad
0\leq k,\lambda <\pi,
\label{eq:root-density-sym}
\end{equation}
as a consequence of the condition \eqref{eq:condition} and the Kronecker deltas in \eqref{eq:initial1-overlapphi}. To prove \eqref{eq:root-density-sym} let us consider the system in a large finite volume $L$. First we note that the rapidities $\tilde{k}_a$ and $k_a$ differ by a fixed constant, so condition \eqref{eq:condition} requires that no two $k_a$ differ by $\pi$.  Since all the eigenstates contributing to the overlap must have $N=L/2$ and no two $k_a$ differing by $\pi$, if there is a hole with $k_a^h$ with $0\leq k_a^h<\pi$, then there must be a rapidity with $k_a=k^h_a-\pi$. Similarly, a hole with $-\pi\leq k_a^h<0$ gives a rapidity with $k_a=k^h_a+\pi$. Close to the thermodynamic limit, this means that for $k\geq0$ the total number of rapidities and holes in an interval $[k,k+\Delta k]$ is equal to the number of rapidities in that interval plus the number of rapidities in the interval $[k-\pi,k-\pi+\Delta k]$
\begin{equation}
L\rhotp(k)\Delta k = L\rhop(k)\Delta k + L\rhop(k-\pi)\Delta k, \qquad 0 \leq k<\pi,
\end{equation}
which gives the above condition on the densities of the rapidities of the particles, since the total density $\rhotp(k)$ is fixed by \eqref{eq:total-rapidity-densities}. A similar argument holds for the rapidities of the spins $\{\lambda_a\}$. Note that \eqref{eq:root-density-sym} fixes the particle and spin densities to be respectively $n=1/2$ and $m=1/4$. This can be seen as follows 
\begin{align}
&n=\int_{-\pi}^{\pi}\!\!\! \mathrm{d}k\, \rhop(k) =\int_{0}^{\pi}\!\!\! \mathrm{d}k\, \left[\rhop(k)+\rhop(k-\pi) \right] =\frac{1}{2}\,, \qquad m=\int_{-\pi}^{\pi}\!\!\! \mathrm{d}k\, \rhos(k) =\int_{0}^{\pi}\!\!\! \mathrm{d}k\, \left[\rhos(k)+\rhos(k-\pi) \right] =\frac{1}{4}\,.
\end{align}

\subsubsection{Reduced Entropy}

Let us now compute the reduced entropy, \emph{i.e.}, the Yang-Yang entropy only taking into account states for which the overlap \eqref{eq:initial1-overlap} is non-zero. In order to do that, it is convenient to re-express the conditions \eqref{eq:root-density-sym} in the following way
\begin{equation}
\rhop^{(1)}(k) + \rhop^{(2)}(k) = \frac{1}{2\pi}, \qquad
\rhos^{(1)}(\lambda) + \rhos^{(2)}(\lambda) = \frac{1}{4\pi},
\label{eq:root-density-sym-2}
\end{equation}
where the new functions are defined on restricted intervals
\begin{subequations}
\begin{align}
\rhop^{(1)}(k) &= \rhop(k), \quad 0\leq k < \pi & \rhos^{(1)}(\lambda) &= \rhos(\lambda), \quad 0\leq \lambda < \pi\\
\rhop^{(2)}(k) &= \rhop(k-\pi), \quad 0\leq k <\pi & \rhos^{(2)}(\lambda) &= \rhos(\lambda-\pi), \quad 0\leq \lambda <\pi.
\end{align}
\label{eq:reduced-root-densities}
\end{subequations}
Having re-expressed the root densities in this form, we can now restrict our attention to the intervals $0\leq k < \pi$ and $0\leq \lambda < \pi$. 

Let us now take the system in a large finite volume $L$ and count the number of eigenstates fulfilling the constraints \eqref{eq:root-density-sym-2}. In the interval $[0,\pi)$ each site can be thought of as filled by either a rapidity of type one or type two. A particle of type one with rapidity $k$ indicates there is a particle with rapidity $k$, whereas a particle of type two with rapidity $k$ indicates there is a particle with rapidity $k-\pi$. The number of states with rapidities in the interval $[k,k+\Delta k]$ is given by the number of ways of arranging $L\rhop^{(1)}(k)\Delta k$ particles of type one and $L\rhop^{(2)}(k)\Delta k$ particles of type two in $L\rhotp(k)\Delta k$ positions. Similarly, the number of states with rapidities in the interval $[\lambda,\lambda+\Delta \lambda]$ is given by the number of ways of arranging $L\rhos^{(1)}(\lambda)\Delta \lambda$ particles of type one and $L\rhos^{(2)}(\lambda)\Delta \lambda$ particles of type two in $L\rhots(\lambda)\Delta \lambda$ positions. Multiplying these two gives the number of states with rapidities in the intervals $[k,k+\Delta k]$ and $[\lambda,\lambda+\Delta \lambda]$
\begin{align}
\exp(\Delta S^\text{red}) &= \frac{[L\rhotp(k)\Delta k]!}{[L\rhop^{(1)}(k)\Delta k]![L\rhop^{(2)}(k)\Delta k]!} \frac{[L\rhots(\lambda)\Delta \lambda]!}{[L\rhos^{(1)}(\lambda)\Delta \lambda]![L\rhos^{(2)}(\lambda)\Delta \lambda]!}\,.
\end{align}
Using Sterling's approximation, taking the logarithm and integrating we have 
\begin{align}
s^\text{red} [\rhop,\rhos] \equiv \lim_{L\rightarrow\infty}\frac{S^\text{red}}{L} =& \int_{0}^\pi \mathrm{d}k \left\{\rhotp(k)\log\rhotp(k)- \rhop^{(1)}(k)\log\rhop^{(1)}(k) - \rhop^{(2)}(k)\log\rhop^{(2)}(k)\right.\nn
&\quad\qquad\left.+ \rhots(k)\log\rhots(k)-\rhos^{(1)}(k)\log\rhos^{(1)}(k) - \rhos^{(2)}(k)\log\rhos^{(2)}(k)\right\}\notag\\
=&  \int_{0}^\pi \mathrm{d}k \left\{s_\text{YY}[\rhop,\rhohp](k)+s_\text{YY}[\rhos,\rhohs](k)\right\}\,,
\label{eq:yangyangentredmu2nu2}
\end{align}
where in the last step we used the definition \eqref{eq:yangyang}. To compare $s^\text{red} [\rhop,\rhos] $ to the Yang-Yang entropy \eqref{eq:yang-yang-entropy}, it is useful to bring it into a different form. Using the symmetry relations \eqref{eq:root-density-sym} we find  
\be
s_\text{YY}[\rhop,\rhohp](k)=s_\text{YY}[\rhop,\rhohp](k-\pi)\,,\qquad\qquad s_\text{YY}[\rhos,\rhohs](k)=s_\text{YY}[\rhos,\rhohs](k-\pi)\,.
\ee
These relations imply 
\begin{align}
s^\text{red} [\rhop,\rhos] = \frac{1}{2} s_{\rm YY} [\rhop,\rhos] \,.
\end{align}
We then see that for the states with $\nu=\mu=2$, the reduced entropy is half of the Yang-Yang entropy. As we will see later, this is not generically true for the states \eqref{eq:initial2-ti}.

\subsubsection{Solution of the Saddle Point Equations}
\label{sec:solutionqdimer}

We now have all the elements required to construct the Quench Action functional, 
\begin{align}
\mathcal F[\rhop, \rhos] &= 2 \mathcal E[\rhop, \rhos]-s^{\text{red}}[\rhop, \rhos]\,.
\label{eq:F_ourcase}
\end{align}
Expressing everything in terms of the root densities \eqref{eq:reduced-root-densities} and using the constraints \eqref{eq:root-density-sym-2}, the saddle point equations can be written as
\be
\frac{\delta\mathcal{F}[ \rhop^{(1)},\rhos^{(1)}]}{\delta  \rhop^{(1)}}\biggl|_{\substack{\rhop^{(1)} = \rhop^{(1)}_s \\ \rhos^{(1)} = \rhos^{(1)}_s}} = 0\,,\qquad\qquad \frac{\delta\mathcal{F}[ \rhop^{(1)},\rhos^{(1)}]}{\delta  \rhos^{(1)}}\biggl|_{\substack{\rhop^{(1)} = \rhop^{(1)}_s \\ \rhos^{(1)} = \rhos^{(1)}_s}} = 0\,.
\label{eq:saddle_point_our_case}
\ee
Taking the variation of the Quench Action functional with respect the root densities $\rhop^{(1)}(k)$ and $\rhos^{(1)}(\lambda)$ and using the constraint  \eqref{eq:root-density-sym-2} we have
\begin{align}
\delta \mathcal{F} &= \int_{0}^\pi\!\! \mathrm{d}k\left\{ 
\left[\log\left(\frac{\rhop^{(1)}(k)}{\rhop^{(2)}(k)} \right)
\!\!-\!\log\left( \frac{1+|q|^2+2|q|\cos( k+\alpha)}{1+|q|^2-2|q|\cos(k+\alpha)}\right)\right] \!\!\delta\rhop^{(1)}(k)\right.\notag\\
&\qquad\qquad+
\left.\left[\log\left(\frac{\rhos^{(1)}(k)}{\rhos^{(2)}(k)} \right)
\!\!-\!\log\left( \frac{1+|p|^2+2|p|\cos(k+\beta)}{1+|p|^2-2|p|\cos(k+\beta)}\right) \right] \!\!\delta\rhos^{(1)}(k)\right\}.
\end{align}
The saddle point equations \eqref{eq:saddle_point_our_case} are then written as
\begin{equation}
\log\left(\frac{\rhop^{(1)}_s(k)}{\rhop^{(2)}_s(k)}\right)=\log\left( \frac{1+|q|^2+2|q|\cos( k+\alpha)}{1+|q|^2-2|q|\cos(k+\alpha)}\right)\,, \qquad
\log\left(\frac{\rhos^{(1)}_s(k)}{\rhos^{(2)}_s(k)}\right)=\log\left( \frac{1+|p|^2+2|p|\cos(k+\beta)}{1+|p|^2-2|p|\cos(k+\beta)}\right)\,.
\label{eq:saddle-point-eqns}
\end{equation}
Solving the saddle point equations along with the constraints \eqref{eq:root-density-sym-2} gives the root densities
\begin{align}
\rhop_s(k)=\frac{1}{4\pi}\left(1+\frac{2|q|\cos(k+\alpha)}{|q|^2+1} \right), \qquad\qquad
\rhos_s(k)=\frac{1}{8\pi}\left(1+\frac{2|p|\cos(k+\beta)}{|p|^2+1} \right)\,.
\label{eq:sp-root-density}
\end{align}
Here, to lighten notation, we do not report the parameter dependence of $\rhop_s(k)$ and $\rhos_s(k)$. To check the validity of our saddle point solution we observe that if the solution \eqref{eq:sp-root-density} is the only relevant saddle point we must have~\cite{PMWK14, mestyan_quenching_2015}
\be
\mathcal F[\rhop_s, \rhos_s]=0\,.
\label{eq:check}
\ee 
This condition ensures that the macrostate \eqref{eq:sp-root-density} encodes all the the relevant information about the initial state in the thermodynamic limit. The validity of \eqref{eq:check} is explicitly shown in Appendix~\ref{app:quench-action-functional-vanishes}. Another non-trivial check on the result \eqref{eq:sp-root-density} is obtained by computing the expectation value of the conserved operators $\hat H_\infty,\hat N$ and $ \hat M$, checking whether they reproduce the initial state values. It is matter of a simple integration to show 
\begin{align}
e_s &= \int_{-\pi}^\pi \!\! {\rm d}k \ \varepsilon(k)\rhop_s(k) = -2\int_{-\pi}^\pi \!\! {\rm d}k \ \cos(k)\rhop_s(k)=-\frac{|q|\cos\alpha}{|q|^2+1}\,,\\
n_s &= \int_{-\pi}^\pi \!\! {\rm d}k \ \rhop_s(k)=\frac{1}{2}\,,
\qquad m_s =\int_{-\pi}^\pi \!\! {\rm d}k \ \rhos_s(k)=\frac{1}{4}\,.
\end{align} 
These values reproduce the initial state averages \eqref{eq:ISdensities} and \eqref{eq:ISenergydensity} for $\nu=\mu=2$ and $a_1=q$.  

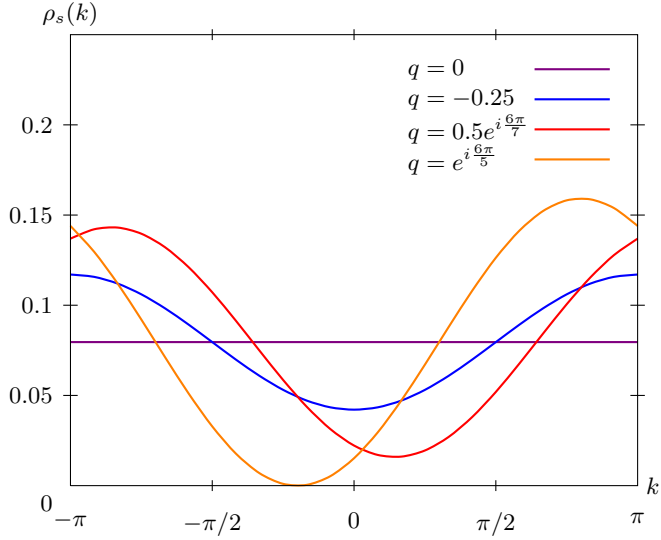
\begin{figure}[t!]
\raisebox{.9mm}{
\begin{tikzpicture}[xscale=1.2, yscale=24]
\draw[scale=1,domain=-3.14159265359:3.14159265359,smooth,solid,variable=\x,violet,  thick] plot ({\x},{0.0795775});
\draw[scale=1,domain=-3.14159265359:3.14159265359,smooth,solid,variable=\x,blue,  thick] plot ({\x},{ 0.0795775*(1.-0.470588*cos(\x r))});
\draw[scale=1,domain=-3.14159265359:3.14159265359,smooth,variable=\x,red,  thick] plot ({\x},{ 0.0795775 *(1.-0.8*cos(\x r-0.449 r))});
\draw[scale=1,domain=-3.14159265359:3.14159265359,smooth,solid,variable=\x,orange,  thick] plot ({\x},{0.0795775*(1.-cos(\x r+0.628 r))});
\draw  node[left] at (-3.24159265359, -0.01) {$0$};
\draw  node[left] at (-3.24159265359, 0.05) {$0.05$};
\draw  node[left] at (-3.24159265359, 0.1) {$0.1$};
\draw  node[left] at (-3.24159265359, 0.15) {$0.15$};
\draw  node[left] at (-3.24159265359, 0.2) {$0.2$};
\draw  node[below] at (0., -0.01) {$0$};
\draw  node[below] at (3.14159265359, -0.01) {$\pi$};
\draw  node[below] at (1.57079632679, -0.01) {$\pi/2$};
\draw  node[below] at (-1.57079632679, -0.01) {$-\pi/2$};
\draw  node[below] at (-3.14159265359, -0.01) {$-\pi$};
\draw (-1.57079632679,-0.003) -- (-1.57079632679, .003);
\draw (0,-0.003) -- (0, .003);
\draw (1.57079632679,-0.003) -- (1.57079632679, .003);
\draw (-1.57079632679,0.247) -- (-1.57079632679, .253);
\draw (0,0.247) -- (0, .253);
\draw (1.57079632679,0.247) -- (1.57079632679, .253);
\draw (-3.22159265359,.2) -- (-3.06159265359,0.2);
\draw (-3.22159265359,.15) -- (-3.06159265359,0.15);
\draw (-3.22159265359,.1) -- (-3.06159265359,0.1);
\draw (-3.22159265359,.05) -- (-3.06159265359,0.05);
\draw (3.22159265359,.2) -- (3.06159265359,0.2);
\draw (3.22159265359,.15) -- (3.06159265359,0.15);
\draw (3.22159265359,.1) -- (3.06159265359,0.1);
\draw (3.22159265359,.05) -- (3.06159265359,0.05);
\draw (-3.14159265359,0) -- (3.14159265359,0) -- (3.14159265359,0.25) -- (-3.14159265359,.25) -- (-3.14159265359,0);
\draw node[right] at (3.14159265359,-0.) {$k$};
\draw (-3.14159265359,.25) node[above] {$\rhop_s(k)$};
\path  (2.,0.185) pic{stationary-density-legend};
\end{tikzpicture}}
\caption{After a quantum quench, the time evolution of observables in the thermodynamic limit is characterised by a set of saddle point root densities $\{\rhop_s(k), \rhos_s(\lambda)\}$. Here we plot the root density of the particles $\rhop_s(k)$ \eqref{eq:sp-root-density} for $q$-dimer initial states with varying $q=0,-0.25,0.5\,e^{i \frac{6\pi}{7}},e^{i \frac{6\pi}{5}}$. The plot of the saddle point root density of the spins $\rhos_s(k)$ is rescaled by a factor of $2$.}
\label{fig:stationary-density}
\end{figure}

From the expressions~\eqref{eq:sp-root-density} we see that the nested-N\'eel state, corresponding to ${q=p=0}$, gives constant root densities. We also see that for real $p\neq0,1$ and $q\neq0,1$ the reflection symmetry, broken by the initial state, is restored in the stationary state. This does not happen when either $p$ or $q$ have a non-zero imaginary part: in this case the stationary distributions are not symmetric around $0$, signalling the presence of local conserved charges which are odd under reflection. It turns out that such asymmetric stationary distributions have interesting effects on the propagation of information (in particular of entanglement) after the quench, an aspect that will be investigated in a future publication~\cite{inpreparation}. Interestingly, even when they are asymmetric, the distributions \eqref{eq:sp-root-density} are obtained by homogeneous boosts of even functions, \emph{i.e.} they are centred around a quasi-momentum $k_0\neq 0$. This is the same as was found in Ref.~\cite{Bucc16} after a quench in the Lieb-Liniger model, starting from a rotating Bose-Einstein condensate. Figure~\ref{fig:stationary-density} shows some examples of the $k$-dependence of $\rhop_s(k)$ for different values of the parameters.

\subsection{Initial states as generalised dimers}

Let us now consider the more general initial states $\ket{\Psi^{\nu\mu}_{\{a_m\}\{b_k\}}}$ and repeat the steps carried out above. 

\subsubsection{Overlaps with the Bethe States}
The overlaps between the states $\ket{\Psi^{\nu\mu}_{\{a_m\}\{b_k\}}}$ and the eigenstates of the Hamiltonian are 
\begin{equation}
\braket{\Psi^{\nu\mu}_{\{a_m\}\{b_k\}}|\Psi_{N,M}(\boldsymbol k;\boldsymbol \lambda)}  = \frac{1}{\sqrt{\nu\mu}} \left(\sum_{j=0}^{\nu\mu-1}e^{ijK}\right)\! \braket{\Phi^{\nu\mu}_{\{a_m\}\{b_k\}}|\Psi_{N,M}(\boldsymbol k;\boldsymbol \lambda)}\,,\qquad K=\sum_{a=1}^{N} k_a\,,
\label{eq:initial2-overlap}
\end{equation}
where the overlap between $\ket{\Phi^{\nu\mu}_{\{a_m\}\{b_k\}}}$ and the states $\ket{\Psi_{N,M}(\boldsymbol k;\boldsymbol \lambda)}$ is given by   
\begin{align}
\braket{\Phi^{\nu\mu}_{\{a_m\}\{b_k\}}|\Psi_{N,M}(\boldsymbol k;\boldsymbol \lambda)}
=&\frac{\delta_{N,{L}/\nu}\delta_{M,{L}/(\nu\mu)}}{L^{N/2} N^{M/2}}
\prod_{j=1}^N\!\left[\frac{\sum_{\ell=0}^{\nu-1}a_\ell e^{-i\ell k_j}}{\sqrt{\sum_{\ell=0}^{\nu-1}|a_\ell|^2}}\right]\!\!
\prod_{j=1}^{M}\!\left[\frac{\sum_{\ell=0}^{\mu-1}b_\ell e^{-i\ell \lambda_j}}{\sqrt{\sum_{\ell=0}^{\mu-1}|b_\ell|^2}}\right]\!
e^{i \Lambda (N+1)/2}\nn
&\times{\det}_{N}\{e^{i \nu  \tilde k_ab} \} {\det}_{M}\{e^{i \mu \lambda_ab} \}\,.
\label{eq:initial2-overlapphi}
\end{align}
Here we again used $\tilde k_a$ as defined in \eqref{eq:ktilde}. Considering the determinants in this expression and reasoning as before we note that, to have a non-zero overlap, no two rapidities $\tilde{k}_a$ can differ by multiples of $2\pi/\nu$ and no two rapidities $\lambda_b$ can differ by multiples $2\pi/\mu$. Combining the conditions with that imposed by requiring the factor in brackets on the r.h.s. of \eqref{eq:initial2-overlap} to be non-zero we have 
\be
\left| \braket{\Psi^{\nu\mu}_{\{0\}\{0\}}|\Psi_{L/\nu,L/(\nu\mu)}(\boldsymbol k;\boldsymbol \lambda)} \right|=
\begin{cases}
C\neq0, &\tilde{k}_a-\tilde k_b\neq0 \!\!\!\mod2\pi/\nu \ \wedge\ \lambda_a-\lambda_b\neq0\!\!\!\mod2\pi/\mu\wedge \ K=0\,\text{mod}\,2\pi\\
0,&\text{otherwise}\,.
\end{cases}
\label{eq:condition2}
\ee 
The constant $C$ can be obtained through an argument similar to the one used in the previous subsection. Here, there are $\nu^{L/\nu}\mu^{L/(\nu\mu)}$ states where no two particle rapidities differ by multiples of $2\pi/\nu$ and no two spin rapidities differ by multiples of $2\pi/\mu$. Of these, $1/(\nu\mu)$ of the states satisfy $K=0\!\!\mod2\pi$, as explained in Appendix~\ref{app:counting-bethe-states}. Therefore, the total number of allowed Bethe states is $\nu^{L/\nu-1}\mu^{L/(\nu\mu)-1}$. Using the fact that the N\'{e}el-like state is normalised, we can write the constant as the reciprocal of the square root of the number of allowed Bethe states
\begin{equation}
C = \frac{1}{\sqrt{\nu^{L/\nu-1}\mu^{L/(\nu\mu)-1}}}\,.
\end{equation}

\subsubsection{Thermodynamic Limit of the Overlaps}
Proceeding as before we  take the thermodynamic limit of the absolute value of the logarithm of \eqref{eq:initial2-overlap} to obtain the first term in the Quench Action functional
\begin{align}
\mathcal{E}[ \rhop,\rhos] &= - \lim_{L\rightarrow\infty}\frac{1}{L}\log |\braket{\Psi^{\nu\mu}_{\{a_m\}\{b_k\}}|\Psi_{\frac{L}{\nu},\frac{L}{\nu\mu}}(\boldsymbol k;\boldsymbol \lambda)}|\notag\\
=&-\frac{1}{2}\int_{-\pi}^\pi \mathrm{d}k
\left\{ \log\left(1+2\sum_{\ell=1}^{\nu-1}|A_\ell|\cos (\ell k+\alpha_\ell)\right)\rhop(k) 
+\log\left(1+2\sum_{\ell=1}^{\mu-1}|B_\ell|\cos (\ell k+\beta_\ell)\right)\rhos(k) \right\}\notag\\
&+\frac{1}{2\nu}\log(\mu^{(1/\mu)}\nu)\,,
\end{align} 
where we have defined
\begin{equation}
A_\ell =\frac{1}{{\sum_{\ell=0}^{\nu-1}|a_\ell|^2}} \sum_{n=\ell}^{\nu-1}{a^*_n  a_{n-\ell}}\,,  \qquad
B_\ell =\frac{1}{{\sum_{\ell=0}^{\mu-1}|b_\ell|^2}} \sum_{n=\ell}^{\mu-1}b^*_n b_{n-\ell}\,, \qquad \alpha_\ell =-\text{arg}\left[A_\ell\right]\,,  \qquad
\beta_\ell =-\text{arg}\left[B_\ell\right]\,,
\label{eq:overlap-coefficients}
\end{equation}
 and we remind the reader that we set $a_0=b_0=1$. In this case, the root densities fulfil
\begin{equation}
\sum_{j=0}^{\nu-1} \rhop\!\left(k-\frac{2\pi}{\nu}j\right) = \frac{1}{2\pi}, \qquad
\sum_{j=0}^{\mu-1} \rhos\!\left(\lambda-\frac{2\pi}{\mu}j\right) = \frac{1}{2\pi \nu}, \qquad
k\in[\pi-\tfrac{2\pi}{\nu},\pi), \ \lambda \in [\pi-\tfrac{2\pi}{\mu},\pi)\,.
\label{eq:conditions2}
\end{equation}
The conditions \eqref{eq:conditions2} can be derived from the constraint \eqref{eq:condition2} and the Kronecker deltas in \eqref{eq:initial2-overlapphi} as follows. Consider the system in a large finite volume $L$. Since all the eigenstates contributing to the overlap must have $N=L/\nu$ and no two ${k}_a$ differing by $2\pi/\nu$, fixing $k\in[\pi-{2\pi}/{\nu},\pi)$ such that 
\be
x(k)=\frac{2\pi}{L} I'_a,\qquad I'_a\in \mathbb Z\,,
\ee
we have two possibilities. The first possibility is that $k$ corresponds to a particle, then all 
\be
k_j\equiv k-2\pi j/\nu\,,
\ee 
for $j=1,\ldots,\nu-1$, correspond to holes. If instead $k$ corresponds to a hole, there is only a single $j$ such that $k_j$ corresponds to a particle. This means that the total number of particles and holes in an interval $\Delta k$ around $k$ is equal to the sum over $j=0,\ldots,\nu-1$ of the number of particles in the intervals of width $\Delta k$ around $k_j$ 
\begin{equation}
L\rhop(k)\Delta k + {L\rho(k-2\pi/\nu)\Delta k} + \ldots+ L\rho(k-2\pi(\nu-1)/\nu)\Delta k=L\rhotp(k) \Delta k\,,
\end{equation}
which yields the first condition of \eqref{eq:conditions2}. A similar argument holds for the rapidities $\lambda$. Note that, as before, the constraints \eqref{eq:conditions2} fix the particle and spin densities to be respectively $n=1/\nu$ and $m=1/(\nu\mu)$. 

\subsubsection{Reduced Entropy}
\label{sec:reducedentropygen}

Let us now write the number of states corresponding to the root densities $\rhop(k)$ and $\rhos(\lambda)$, subject to the constraints \eqref{eq:conditions2}, which we rewrite as
\be
\sum_{j=1}^\nu \rhop^{(j)}(k) = \frac{1}{2\pi},\qquad\qquad
\sum_{j=1}^\mu \rhos^{(j)}(\lambda) = \frac{1}{2\pi \nu}\,,
\ee
where we introduced 
\begin{align}
&\rhop^{(j)}(k) \equiv \rhop(k-\tfrac{2\pi}{\nu}(j-1))\,,\qquad \pi-\frac{2\pi}{\nu}\leq k < \pi\,,\qquad j=1.\ldots,\nu\,,\notag\\
&\rhos^{(j)}(\lambda) \equiv \rhos(\lambda-\tfrac{2\pi}{\mu}(j-1))\,,\qquad  \pi-\frac{2\pi}{\mu}\leq \lambda < \pi\,,\qquad j=1,\ldots,\mu\,.
\label{eq:root-densities-initial2}
\end{align}
This allows us to easily count the number of possible states in the intervals $[k,k+\Delta k]$ and $[\lambda,\lambda + \Delta \lambda]$, by restricting our attention to   $k\in[\pi-2\pi/\nu,\pi)$ and $\lambda\in[\pi-2\pi/\mu,\pi)$ and considering $\nu$ different types of particle rapidities and $\mu$ different types of spin rapidities, respectively. The exponential of the reduced entropy is therefore given by
\begin{align}
\exp(\Delta S^\text{red}) &= \frac{[L\rhotp(k)\Delta k]!}{\prod_{j=1}^\nu [L\rhop^{(j)}(k)\Delta k]!} \frac{[L\rhots(\lambda)\Delta \lambda]!}{\prod_{j=1}^\mu [L\rhos^{(j)}(\lambda)\Delta \lambda]!}\,,
\end{align}
from which we find the reduced entropy by taking the logarithm and using Sterling's approximation for large $L$ 
\begin{align}
s^\text{red}[\rhop,\rhos] =& \int_{\pi-\frac{2\pi}{\nu}}^\pi \mathrm{d}k \left\{\rhotp(k)\log\rhotp(k) - \sum_{j=1}^\nu\rhop^{(j)}(k)\log\rhop^{(j)}(k)\right\}\notag\\
&+\int_{\pi-\frac{2\pi}{\mu}}^\pi \mathrm{d}\lambda \left\{\rhots(\lambda)\log\rhots(\lambda) -\sum_{j=1}^\mu\rhos^{(j)}(\lambda)\log\rhos^{(j)}(\lambda)\right\}.
\label{eq:yangyangreduced}
\end{align}
A convenient rewriting of \eqref{eq:yangyangreduced} reads as 
\begin{align}
s^\text{red}[\rhop,\rhos] =& \int_{\pi-\frac{2\pi}{\nu}}^\pi \!\!\mathrm{d}k\,\rhotp(k)\log\rhotp(k) - \int_{-\pi}^\pi \mathrm{d}k\,\, \rhop(k)\log\rhop(k)+\int_{\pi-\frac{2\pi}{\mu}}^\pi\!\!\mathrm{d}\lambda\,\, \rhots(\lambda)\log\rhots(\lambda) -\int_{-\pi}^\pi\!\!\mathrm{d}\lambda\,\,\rhos(\lambda)\log\rhos(\lambda)\,.
\label{eq:yangyangreduced2}
\end{align}
This expression, as opposed to the $\nu=\mu=2$ case, is not generically proportional to the Yang-Yang entropy $s_\text{YY}[\rhop,\rhos]$. This can be, \emph{e.g.}, seen by noting 
\be
\frac{\rm d}{{\rm d}a}\left(\frac{s^\text{red}[a,0] }{s_\text{YY}[a,0] }\right)\neq0\qquad\qquad a\in\left[0,\frac{1}{2\pi}\right]\,.
\ee

\subsubsection{Solution of the Saddle Point Equations}

The quench action functional is constructed as before (\emph{cf}. \eqref{eq:F_ourcase}). Writing everything in terms of the root densities $\rhop^{(j)}(k)$ and $\rhos^{(j)}(\lambda)$ defined in Eq.~\eqref{eq:root-densities-initial2}, the saddle point equations read as
\be
\frac{\delta\mathcal{F}[\{ \rhop^{(i)},\rhos^{(j)}\}]}{\delta  \rhop^{(i)}}\biggl|_{\substack{\rhop^{(i)} = \rhop^{(i)}_s \\ \rhos^{(j)} = \rhos^{(j)}_s}} = 0\,,\qquad \frac{\delta\mathcal{F}[ \{\rhop^{(i)},\rhos^{(j)}\}]}{\delta  \rhos^{(j)}}\biggl|_{\substack{\rhop^{(i)} = \rhop^{(i)}_s \\ \rhos^{(j)} = \rhos^{(j)}_s}} = 0\,,\qquad i=1,\ldots,\nu-1,\qquad j=1,\ldots,\mu-1\,.
\label{eq:saddle_point_our_case_gen}
\ee
Here to have independent functions we write $\rhop^{(\nu)}(k)$ and $\rhos^{(\mu)}(\lambda)$ in terms of the other root densities using the conditions \eqref{eq:conditions2}, giving $\mu+\nu-2$ equations. The variation of the functional $\mathcal F[\{ \rhop^{(i)},\rhos^{(j)}\}]$ is written as 
\begin{align}
\delta F =
&-\int_{\pi-\frac{2\pi}{\nu}}^\pi \mathrm{d}k 
\sum_{j=1}^{\nu-1}\left[\log\left(\frac{\rhop^{(j)}(k)}{\rhop^{(\nu)}(k)}\right)
-\log\left(\frac{1+2\sum_{\ell=1}^{\nu-1}|A_\ell|\cos (\ell k-\tfrac{2\pi\ell}{\nu}(j-1)+\alpha_\ell)}{1+2\sum_{\ell=1}^{\nu-1}|A_\ell|\cos(\ell k+\tfrac{2\pi\ell}{\nu}+\alpha_\ell)}\right) 
\right]\delta\rhop^{(j)}(k) \nn
&-\int_{\pi-\frac{2\pi}{\mu}}^\pi \mathrm{d}\lambda
\sum_{j=1}^{\mu-1}\left[\log\left(\frac{\rhos^{(j)}(\lambda)}{\rhos^{(\mu)}(\lambda)}\right)
- \log\left(\frac{1+2\sum_{\ell=1}^{\mu-1}|B_\ell|\cos(\ell \lambda-\tfrac{2\pi\ell}{\mu}(j-1)+\beta_\ell)}{1+2\sum_{\ell=1}^{\mu-1}|B_\ell|\cos(\ell\lambda+\tfrac{2\pi\ell}{\mu}+\beta_\ell)}\right)\right]\delta\rhos^{(j)}(\lambda)\,.
\end{align}
As in the case $\mu=\nu=2$, the chemical potential term does not depend on the root densities and therefore vanishes when taking the variation. The saddle point equations obtained by setting the variation of the Quench Action functional to zero then read as
\begin{subequations}
\begin{align}\log\left(\frac{\rhop^{(j)}_s(k)}{\rhop^{(\nu)}_s(k)}\right)
&=\log\left(\frac{1+2\sum_{\ell=1}^{\nu-1}|A_\ell|\cos (\ell k-\tfrac{2\pi\ell}{\nu}(j-1)+\alpha_\ell)}{1+2\sum_{\ell=1}^{\nu-1}|A_\ell|\cos(\ell k+\tfrac{2\pi\ell}{\nu}+\alpha_\ell)}\right) \,, & j&=1,\ldots,\nu-1\,,\\
\log\left(\frac{\rhos^{(j)}_s(\lambda)}{\rhos^{(\mu)}_s(\lambda)}\right) 
&=\log\left(\frac{1+2\sum_{\ell=1}^{\mu-1}|B_\ell|\cos(\ell \lambda-\tfrac{2\pi\ell}{\mu}(j-1)+\beta_\ell)}{1+2\sum_{\ell=1}^{\mu-1}|B_\ell|\cos(\ell\lambda+\tfrac{2\pi\ell}{\mu}+\beta_\ell)}\right)\,, & j&=1,\ldots,\mu-1\,.
\end{align}
\end{subequations}
The solution reads as
\begin{align}
\rhop_s(k) = \frac{1}{2\pi\nu}\left(1+2\sum_{\ell=1}^{\nu-1}|A_\ell|\cos (\ell k+\alpha_\ell)\right)\!,\qquad
\rhos_s(\lambda) = \frac{1}{2\pi\nu \mu}\left(1+2\sum_{\ell=1}^{\mu-1}|B_\ell|\cos(\ell \lambda+\beta_\ell)\right)\,,
\label{eq:result_root_densities}
\end{align}
where the coefficients $A_\ell$, $B_\ell$, $\alpha_\ell$ and $\beta_\ell$ are defined in Eq.~\eqref{eq:overlap-coefficients}. Also in this case the Quench Action functional vanishes when evaluated at the saddle point root densities, as shown in Appendix~\ref{app:quench-action-functional-vanishes}. The expectation values of the conserved operators $\hat H_\infty,\hat N$ and $\hat M$ read as
\begin{align}
e_s &= \int_{-\pi}^\pi \!\! {\rm d}k \ \varepsilon(k)\rhop_s(k) = -2\int_{-\pi}^\pi \!\! {\rm d}k \ \cos(k)\rhop_s(k)=-\frac{2}{\nu} |A_1| \cos\alpha_1\,,\\
n_s &= \int_{-\pi}^\pi \!\! {\rm d}k \ \rhop_s(k)=\frac{1}{\nu}\,,
\qquad m_s =\int_{-\pi}^\pi \!\! {\rm d}k \ \rhos_s(k)=\frac{1}{\nu\mu}\,,
\end{align} 
reproducing the initial state averages \eqref{eq:ISdensities} and \eqref{eq:ISenergydensity}.

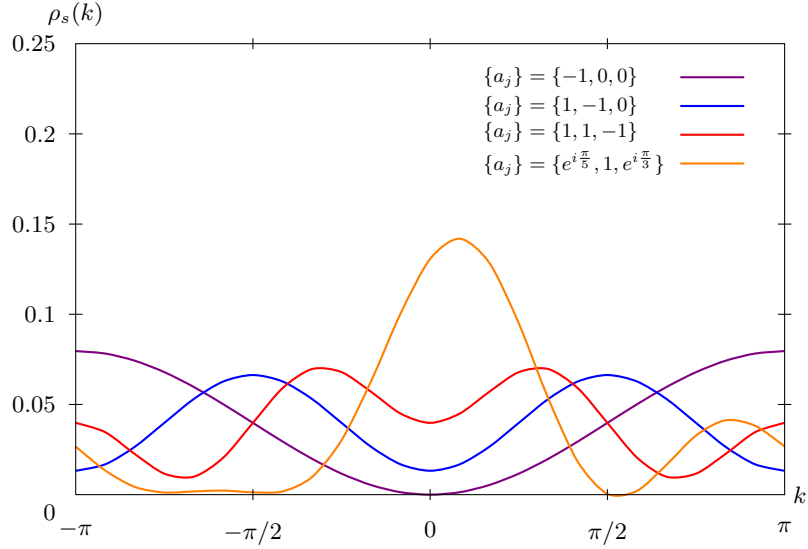
\begin{figure}[t!]
\raisebox{.9mm}{
\begin{tikzpicture}[xscale=1.5, yscale=24]
\draw[scale=1,domain=-3.14159265359:3.14159265359,smooth,solid,variable=\x,violet,  thick] plot ({\x},{(0.0397887*(1. -1.*cos(\x r)))});\draw[scale=1,domain=-3.14159265359:3.14159265359,smooth,solid,variable=\x,blue,  thick] plot ({\x},{ 0.0397887*(1.-0.666667*cos (2.*\x r))});
\draw[scale=1,domain=-3.14159265359:3.14159265359,smooth,variable=\x,red,  thick] plot ({\x},{ 0.0397887*(2.*(0.25*cos (\x r)-0.25*cos (3.*\x r))+1.)});
\draw[scale=1,domain=-3.14159265359:3.14159265359,smooth,solid,variable=\x,orange,  thick] plot ({\x},{0.0397887*(2.*(0.57*cos (\x r-0.38 r)+0.5*cos (2.*\x r-0.21 r)+0.25*cos (3.*\x r-1.05 r))+1.)});
\draw  node[left] at (-3.24159265359, -0.01) {$0$};
\draw  node[left] at (-3.24159265359, 0.05) {$0.05$};
\draw  node[left] at (-3.24159265359, 0.1) {$0.1$};
\draw  node[left] at (-3.24159265359, 0.15) {$0.15$};
\draw  node[left] at (-3.24159265359, 0.2) {$0.2$};
\draw  node[left] at (-3.24159265359, 0.25) {$0.25$};
\draw  node[below] at (0., -0.01) {$0$};
\draw  node[below] at (3.14159265359, -0.01) {$\pi$};
\draw  node[below] at (1.57079632679, -0.01) {$\pi/2$};
\draw  node[below] at (-1.57079632679, -0.01) {$-\pi/2$};
\draw  node[below] at (-3.14159265359, -0.01) {$-\pi$};
\draw (-1.57079632679,-0.003) -- (-1.57079632679, .003);
\draw (0,-0.003) -- (0, .003);
\draw (1.57079632679,-0.003) -- (1.57079632679, .003);
\draw (-1.57079632679,0.247) -- (-1.57079632679, .253);
\draw (0,0.247) -- (0, .253);
\draw (1.57079632679,0.247) -- (1.57079632679, .253);
\draw (-3.20159265359,.2) -- (-3.08159265359,0.2);
\draw (-3.20159265359,.15) -- (-3.08159265359,0.15);
\draw (-3.20159265359,.1) -- (-3.08159265359,0.1);
\draw (-3.20159265359,.05) -- (-3.08159265359,0.05);
\draw (3.20159265359,.2) -- (3.08159265359,0.2);
\draw (3.20159265359,.15) -- (3.08159265359,0.15);
\draw (3.20159265359,.1) -- (3.08159265359,0.1);
\draw (3.20159265359,.05) -- (3.08159265359,0.05);
\draw (-3.14159265359,0) -- (3.14159265359,0) -- (3.14159265359,0.25) -- (-3.14159265359,.25) -- (-3.14159265359,0);
\draw node[right] at (3.14159265359,-0.) {$k$};
\draw (-3.14159265359,.255) node[above] {$\rhop_s(k)$};
\node at (1.6,.207) {
	\begin{tikzpicture}[scale=0.85, every node/.style={scale=0.85}]
		\path  (0,0) pic{stationary-density-legend-2};
	\end{tikzpicture}
};
\end{tikzpicture}}
\caption{When the initial state of the system is a generalised dimer~\eqref{eq:initial2-ti}, the saddle point root densities $\{\rhop_s(k), \rhos_s(\lambda)\}$ are sums of cosine functions with varying period, instead of a single cosine function in the case of the $q$-dimer. This figure shows some examples of root densities of the particles $\rhop_s(k)$ \eqref{eq:result_root_densities} with $\nu=4$.
The saddle point root density of the spins $\rhos_s(k)$ is again a simple rescaling of the one of the particles.
}
\label{fig:stationary-density-2}
\end{figure}

Even though the solutions \eqref{eq:result_root_densities} have a more complicated structure compared to \eqref{eq:sp-root-density}, we again observe a restoration of reflection symmetry when the initial states have real coefficients. When the coefficients are complex, however, we see that \eqref{eq:result_root_densities} are not generically simple homogeneous boosts of symmetric quasi-momentum distributions: they have a genuinely non-symmetric structure. Some examples of the $k$ dependence of $\rhop_s(k)$ for different values of the parameters are reported in Figure~\ref{fig:stationary-density-2}.

\section{Conclusions}
\label{sec:conclusions}

In this paper we have studied quantum quenches in the Hubbard model in the limit of infinite interaction. We considered a large family of initial states constructed as follows. First one takes the ground state of a locally interacting Hamiltonian, invariant under translations of a number of lattice sites strictly greater than one. Then one constructs a translationally invariant linear combination, which is amenable to a simpler analysis. The class of initial states considered includes for example the (translationally invariant) ``nested N\'eel state", where the fermions occupy one site every two and one fermion every two has spin down. It also includes its generalisations to arbitrary densities of particles and spin. The initial states considered are interesting for two main reasons. Firstly, their experimental realisation appears well within the bounds of possibility. Secondly, due to their simple structure, they are the natural candidates for studying quenches to the Hubbard model with finite interaction. To make our treatment easily generalisable to interacting models we used the Quench Action method to analytically determine the stationary state root densities. Generic states in the family we consider are not reflection symmetric.  Among non-reflection symmetric states we identified two subclasses: evolving from states in the first subclass the reflection symmetry is restored in the stationary state while evolving from states in the second class the symmetry remains broken. In the latter case, the stationary rapidity distributions cannot generically be written as symmetric distributions centred around a momentum $k_0\neq0$ but bear a genuinely asymmetric structure. The effects of such an asymmetric structure on the ratio between diagonal and thermodynamic entropies (see  Ref.~\cite{CA17}) and on the entanglement dynamics are currently under investigation~\cite{inpreparation}. 

Since we studied the problem for infinite interaction $U=\infty$, we were able use the simple XX basis for the ``spin sector" (\emph{cf}. Section~\ref{sec:eigenstates}). A further step towards the understanding of the fully interacting Hubbard model is to use the XXX basis for the spin sector. This basis appears naturally taking the ${U\rightarrow \infty}$ limit of the Bethe states of the finite $U$ Hubbard model and encodes the first $t/U$ correction to our result~\cite{os:strongcoupling}. A subset of the initial states considered in this work remains exactly treatable also using the XXX basis for the spin sector; this problem is also currently under investigation.

Other interesting directions for future research are suggested by the simple structure of the model studied here. Firstly, one can compute correlation functions on the saddle point state, generalising the thermal state results of Refs.~\cite{IPA:j0model, IP:twocomponentgases} to generic stationary states. This model is also a good candidate for performing analytic calculations of the full time evolution of relevant expectation values, \emph{e.g.}, the one-body density matrix. To do that, there are two main routes that can be considered. One is using the mapping~\cite{K:Hubbardmapping} to a free fermionic theory, while the other is adopting the Quench Action approach for finite times as in Refs.~\cite{dc-14, PC:anyons}. 
Finally, it would be very interesting to study transport problems in the model examined here, as its simple structure allows for a fully analytical treatment of the nested case.

\section{Acknowledgements}
We are grateful to Fabian Essler for drawing our attention to Refs.~\cite{K:Hubbardmapping, K:Operators} and for stimulating discussions.  We sincerely thank Lorenzo Piroli and Maurizio Fagotti for useful suggestions and valuable comments. BB and PC acknowledge the financial support by the ERC under Starting Grant 279391 EDEQS.

\appendix
\section{Number of allowed Bethe States}
\label{app:counting-bethe-states}
Here we calculate the number of Bethe states \eqref{eq:wavefunction} that have non-zero overlap with states in the class \eqref{eq:initial2-ti}. These Bethe states satisfy condition \eqref{eq:condition2}: no two particle rapidities can differ by integer multiples of $2\pi/\nu$, no two spin rapidities by integer multiples of $2\pi/\mu$, and the total momentum $K$ must be zero modulo $2\pi$. 

We begin by constructing one state satisfying the first constraint: 
the one where the particle rapidities $\tilde k$ (\emph{cf}. Eq.~\eqref{eq:ktilde}) take the $N$ values  $\tilde{k}'_a\in[\pi-\tfrac{2\pi}{\nu},\pi)$ and the spin rapidities the $M$ values $\lambda'_a\in[\pi-\tfrac{2\pi}{\mu},\pi)$
\begin{subequations}
\begin{align}
\tilde{k}'_a &= \frac{2\pi}{L}\left(L-\left\lfloor \frac{L}{2}\right\rfloor-N+a\right), & a&=0,1,\ldots,N-1\,,\\
\lambda'_b &= \frac{2\pi}{N}\left(N-\left\lfloor \frac{N}{2}\right\rfloor+\frac{M+1\!\!\mod2}{2}-M+b\right), & b&=0,1,\ldots,M-1\,,
\end{align}
\end{subequations}
where $L=\nu N = \nu \mu M$. All other Bethe states satisfying the first constraint can be constructed from this one by shifting the rapidities of the particles by integer multiples of $2\pi/\nu$ and those of the spins by integer multiples of $2\pi/\mu$
\begin{align}
(\tilde{k}_0,\tilde{k}_1,\ldots,\tilde{k}_{N-1}) &= 
(\tilde{k}'_0-\tfrac{2\pi}{\nu}r_0,\tilde{k}'_1-\tfrac{2\pi}{\nu}r_1,\ldots,\tilde{k}'_{N-1}-\tfrac{2\pi}{\nu}r_{N-1})\,, 
& r_i &=0,1,\ldots,\nu-1\,,\notag\\
(\lambda_0,\lambda_1,\ldots,\lambda_{M-1}) &= 
(\lambda'_0-\tfrac{2\pi}{\mu}s_0,\lambda'_1-\tfrac{2\pi}{\mu}s_1,\ldots,\lambda'_{M-1}-\tfrac{2\pi}{\mu}s_{M-1})\,,
& s_i &=0,1,\ldots,\mu-1\,.
\label{eq:explicit-rapidities}
\end{align}
There are $\nu^N\mu^M$ such states, however, they include those which do not satisfy the second condition: that the total momentum must be zero modulo $2\pi$. We argue that $1/(\nu\mu)$ of the states \eqref{eq:explicit-rapidities} satisfy this condition as follows. 

We partition the full set of rapidities \eqref{eq:explicit-rapidities} into subsets containing $\nu\mu$ states where each subset takes all possible values for $(r_0,s_0)$ and fixes the rest. We denote the total momentum of each set of rapidities by 
\begin{equation}
K_{r_0,s_0} =\left(K_{0,0} -\frac{2\pi}{\nu} r_0 - \frac{2\pi}{\nu\mu} s_0 \right)\!\!\mod2\pi\,,
\label{eq:subset-tot-momentum}
\end{equation}
where $K_{0,0}$ is the momentum of the set for $s_0=r_0=0$. In the next subsection we show that the total momentum of a state, for any set of rapidities \eqref{eq:explicit-rapidities}, can always be written as $2\pi \gamma /(\nu\mu)$ for some $\gamma=0,1,\ldots,\nu\mu-1$. Using this in \eqref{eq:subset-tot-momentum} for $K_{0,0}$ we find
\begin{equation}
K_{r_0,s_0} =\left(\frac{2\pi\gamma}{\nu\mu}-\frac{2\pi}{\nu} r_0 - \frac{2\pi}{\nu\mu} s_0 \right)\!\!\mod2\pi\,.
\end{equation}
Using this expression, we find that the equation $K_{r_0,s_0} =0\!\!\mod2\pi$ has the unique solution
\begin{equation}
s_0 = \gamma \mod \mu, \qquad r_0 = s_0-\gamma \mod\nu\,.
\end{equation}
This means, each group of $\nu\mu$ states has exactly one with total momentum zero modulo $2\pi$. Therefore, of the total $\nu^N\mu^M$ states represented in \eqref{eq:explicit-rapidities}, $1/(\nu\mu)$ of them have momentum zero, and the total number of allowed Bethe states is $\nu^{N-1}\mu^{M-1} = \nu^{L/\nu-1}\mu^{L/(\nu\mu)-1}$.

\subsection{Total momentum of Bethe states \eqref{eq:explicit-rapidities}}
The total momentum of a generic Bethe state can be written as 
\begin{align}
K = \left(\widetilde{K}+\frac{\Lambda\!\!\!\!\mod2\pi}{\nu}\right)\!\!\!\!\mod2\pi\,,  \qquad\qquad \widetilde{K} =  \sum_{a=0}^{N-1}\tilde k_a\,,  \qquad\qquad \Lambda = \sum_{b=0}^{M-1}\lambda_b\,. 
\end{align}
Considering the Bethe states of the form \eqref{eq:explicit-rapidities} we then have 
\begin{subequations}
\begin{align}
\widetilde{K} &= \frac{\pi}{\nu}\left[2\left(L-\left\lfloor \frac{L}{2} \right\rfloor - \sum_{a=0}^{N-1}r_a  \right) -(N+1)\right]
= \begin{cases}
\frac{2\alpha\pi}{\nu}, & N \text{ odd,}\\[2pt]
\frac{(2\alpha+1)\pi}{\nu}, & N \text{ even,}
\end{cases}
& \alpha &= 0,1,\ldots, \nu-1\\
\Lambda &= \frac{2\pi}{\mu}\left(N-\left\lfloor \frac{N}{2} \right\rfloor -\left\lfloor \frac{M}{2} \right\rfloor -1 - \sum_{b=0}^{M-1}s_b \right)
= \frac{2\beta\pi}{\mu}, & \beta &= 0,1,\ldots,\mu-1,\\
K &= \widetilde{K}+\frac{\Lambda\!\!\!\!\mod2\pi}{\nu}
= \begin{cases}
\frac{2(\alpha\mu+\beta)\pi}{\nu\mu}, & N \text{ odd,}\\[2pt]
\frac{((2\alpha+1)\mu+2\beta)\pi}{\nu\mu}, & N \text{ even,}
\end{cases}
\end{align}
\end{subequations}
where all the equalities hold modulo $2\pi$. We see that for $N$ odd, the total momentum $K$ takes values $K=2\pi\gamma/(\nu\mu)$, $\gamma=0,1,\ldots,\nu\mu-1$. In the case $N$ even this is not obvious. We recall, however, that for our initial states \eqref{eq:initial2-ti} at least one of the following properties hold: (i) $\mu$ even, (ii) $M$ odd. If $\mu$ is even, we get the same values of $K$ as in the case $N$ odd. If $M$ is odd and $N$ is even, it must be that $\mu$ is again even, since $N =\mu M$, giving the same possible values for $K$.

\section{Consistency check on the saddle point solution}
\label{app:quench-action-functional-vanishes}

In the body of this paper we calculate the saddle point root densities $\boldsymbol{\rho_s}$ for various initial states by minimising the Quench Action functional $\mathcal{F}[\boldsymbol{\rho}]$. Since the initial states are normalised to one, if the saddle point root densities encode all the information on the initial state in the thermodynamic limit we must have~\cite{PMWK14, mestyan_quenching_2015}
\be
\mathcal F[\boldsymbol \rho_s]=0\,.
\ee 
In this appendix, we perform this check on our solutions. We immediately consider the most general case, where the initial state is a generalised dimer \eqref{eq:initial2-ti}; the case \eqref{eq:sp-root-density} is recovered by setting $\mu=\nu=2$. The Quench Action functional evaluated at the saddle point root densities can be written as follows
\begin{equation}
\mathcal{F}[\rhop_s,\rhos_s] = 2\mathcal{E}[\rhop_s,\rhos_s]-s^\text{red}[\rhop_s,\rhos_s]\,.
\end{equation}
Expressing the saddle point root densities in terms of the reduced root densities \eqref{eq:root-densities-initial2} we find 
\begin{align}
2\mathcal{E}[\rhop_s,\rhos_s] &=
-\int_{-\pi}^\pi \mathrm{d}k\left[\log(2\pi\nu\rhop_s(k))\rhop_s(k) + \log(2\pi\mu\rhos_s(k))\rhos_s(k) \right]
+\frac{1}{\nu}\log(\mu^{(1/\mu)}\nu)\nn
&= -n\log(2\pi\nu) - m\log\left(\frac{2\pi\mu}{n} \right)
- \int_{\pi-\tfrac{2\pi}{\nu}}^\pi \mathrm{d}k\ \sum_{j=1}^\nu \rhop_s^{(j)}(k)\log\rhop^{(j)}_s(k)\nn
&- \int_{\pi-\tfrac{2\pi}{\mu}}^\pi \mathrm{d}k\ \sum_{j=1}^\mu \rhos_s^{(j)}(k)\log\rhos^{(j)}_s(k)
+\frac{1}{\nu}\log(\mu^{(1/\mu)}\nu)\,,
\end{align}
where
\begin{subequations}
\begin{align}
&\rhop_s^{(j)}(k) \equiv \rhop_s(k-\tfrac{2\pi}{\nu}(j-1))\,,\qquad \pi-\frac{2\pi}{\nu}\leq k < \pi\,,\qquad j=1.\ldots,\nu\,,\\
&\rhos_s^{(j)}(\lambda) \equiv \rhos_s(\lambda-\tfrac{2\pi}{\mu}(j-1))\,,\qquad  \pi-\frac{2\pi}{\mu}\leq \lambda < \pi\,,\qquad j=1,\ldots,\mu\,.
\end{align}
\end{subequations}
The reduced entropy instead reads as 
\begin{align}
s^\text{red}[\rhop_s,\rhos_s] =&
 \int_{\pi-\tfrac{2\pi}{\nu}}^\pi \mathrm{d}k\left[ \frac{1}{2\pi}\log\left(\frac{1}{2\pi} \right) -  \sum_{j=1}^\nu \rhop_s^{(j)}(k)\log\rhop^{(j)}_s(k)\right]\notag\\
 & + \int_{\pi-\tfrac{2\pi}{\mu}}^\pi \mathrm{d}k\ \left[ \frac{n}{2\pi}\log\left(\frac{n}{2\pi} \right)- \sum_{j=1}^\mu \rhos_s^{(j)}(k)\log\rhos^{(j)}_s(k)\right]\!.
\end{align}
Adding these terms together, all the integrals over the reduced root densities cancel, leaving
\begin{equation}
\mathcal{F}[\rhop_s,\rhos_s] = 
-n\log(2\pi\nu) - m\log\left(\frac{2\pi\mu}{n} \right)
+\frac{1}{\nu}\log(\mu^{(1/\mu)}\nu)
-\int_{\pi-\tfrac{2\pi}{\nu}}^\pi \mathrm{d}k\ \frac{1}{2\pi}\log\left(\frac{1}{2\pi} \right)
-\int_{\pi-\tfrac{2\pi}{\mu}}^\pi \mathrm{d}k\ \frac{n}{2\pi}\log\left(\frac{n}{2\pi} \right)
=0\,,
\end{equation}
where we have used $n=1/\nu$ and $m=1/(\mu\nu)$.

\end{document}